\documentclass[]{aa}  

\usepackage{natbib}
\bibpunct{(}{)}{;}{a}{}{,}
\usepackage{graphicx}
\usepackage{revsymb}	
\usepackage{amsmath}	
\usepackage[usenames]{color}	
\usepackage{epstopdf}	
\usepackage{svg}
\usepackage{txfonts}
\usepackage{amssymb}
\usepackage{color}

\usepackage{geometry} 
\usepackage[parfill]{parskip} 
\usepackage{amssymb}
\usepackage{slantsc}
\usepackage{array}
\usepackage{url}

\usepackage{subfloat}
\usepackage{ float}

\usepackage{amsfonts}
\usepackage[colorlinks=true,linkcolor=blue, urlcolor=blue, citecolor=blue]{hyperref}
\usepackage{sidecap}

\usepackage{multicol}
\usepackage{graphicx}
\usepackage{xcolor}

\usepackage{nicefrac}
\usepackage{subcaption}
\usepackage{epsfig}
\usepackage{soul}
\usepackage{enumitem}
\usepackage{leftidx}
\usepackage{lineno}

\colorlet{rouge}{red!70!darkgray}

\begin{document}

\title{Detailed theoretical analysis of core Helium-burning stars:\\ Mixed mode patterns   }
\subtitle{I. Impact of the He-flash discontinuity and of induced semi-convection}
\author{L. Panier\inst{1} \and G. Buldgen\inst{1} \and M. Matteuzzi\inst{2,3}  \and R. Scuflaire\inst{1} \and M.A Dupret\inst{1}  \and A. Noels\inst{1} \and A. Miglio\inst{2,3} }
\institute{STAR Institute, Université de Liège, Liège, Belgium \and Department of Physics \& Astronomy "Augusto Righi", University of Bologna, via Gobetti 93/2, 40129, Bologna, Italy \and INAF-Astrophysics and Space Science Observatory of Bologna, via Gobetti 93/3, 40129, Bologna, Italy}
\abstract{Recent space missions such as CoRoT, \textit{Kepler}, and TESS have made asteroseismology a powerful tool for studying the internal structure of stars. Red giants, in particular, are central in these studies due to their rich oscillation spectra, which provide details of both their core and envelope through their mixed oscillation modes. Despite these advances, models of core helium-burning red giants struggle to reproduce the observed oscillation spectra, particularly because of uncertainties in the treatment of mixing processes such as overshooting and semi-convection. These discrepancies highlight the need for further asteroseismic constraints to improve stellar models.}
{We aim to identify the key structural features influencing asteroseismic observations and how seismic signatures relate to  internal chemical composition by investigating the asteroseismic properties of core helium-burning stars with a consistent treatment of induced semi-convection and overshooting.}
{We use a new version of the Liège stellar evolution code and the Liège adiabatic oscillation code to compute and analyse the mixed-mode oscillation patterns of various models of core helium-burning stars.}
{We find that sharp transitions in the chemical composition and overshooting in the central part of the models significantly affect the mixed-mode oscillation spectra of core helium-burning stars. Overshooting variations alter the size of the semi-convective zone, which  locally modifies the Brunt–Väisälä frequency and, thereby, the observed period spacing. Notably, our models indicate that modifications in overshooting are balanced by adjustments in the semi-convective layers, maintaining a consistent total mixed-core size across models. Thus, stellar evolution is minimally affected by these internal adjustments, unlike the seismic signatures, as seen in the Brunt–Väisälä frequency profile. The semi-convective zone also introduces additional seismic trapping, although for advanced models, modes confined within this region are unlikely to be detectable due to their high energy density and minimal impact on surrounding modes.}{We highlight the importance of detailed seismic studies to characterise mixing processes near the convective core in core helium-burning stars. This provides a first step towards constraining the chemical composition gradient. In addition, we notice a balance between the extension of overshooting and semi-convective zones, affecting the oscillation spectra.}
\keywords{asteroseismology – stars: interiors – stars: evolution}
\maketitle

\section{Introduction}

In recent decades, high-quality data from space missions such as CoRoT \citep{auvergne_corot_2009}, \textit{Kepler} \citep{borucki_kepler_2010}, and TESS \citep{ricker_transiting_2014} have established asteroseismology as a powerful tool for probing the internal structure and dynamics of stars. It {has been particularly efficient in analysing the rich seismic spectra of red giant (RGB) stars }\citep[see e.g.][and references therein]{Chaplin2013,Hekker2017}.  {In RGBs, pressure modes (p modes) are acoustic waves that mainly propagate  in the extended convective envelope, where pressure is the dominant restoring force. Gravity modes (g modes), on the other hand, are buoyancy-driven and propagate in the dense radiative core. Because RGBs have both a compact core and an extended envelope, some oscillations become mixed modes, behaving as g modes in the core and p modes in the envelope, which allows us to probe both regions. }
Red giant stars serve as laboratories for testing the theory of stellar structure and evolution, exposing several issues with theoretical models \citep{siess_thermohaline_2009,lagarde_thermohaline_2011,mosser_spin_2012,deheuvels_seismic_2014,pincon_probing_2020}. 
One major achievement of their seismic study has been the ability to distinguish between different evolutionary stages of RGBs, namely hydrogen-shell burning and core-helium burning (CHeB) stars \citep{montalban_seismic_2010,bedding_gravity_2011,mosser_universal_2011}. This distinction was made possible by the difference in the density contrast between the core and the average density. The onset of nuclear reactions reduces the central density in a CHeB star by almost a factor of 10 compared to that of a RGB star. This shifts the region that contributes the most to the Brunt–Väisälä frequency to larger radii, with a clear effect on the value of the period spacing \citep{montalban_testing_2013}. Moreover, this different density contrast modifies the coupling between the g- and p-mode cavities, affecting -among other things- the regularity of the spectrum \citep{montalban_seismic_2010}.

Furthermore, a significant discrepancy of two orders of magnitude was revealed between the predicted core rotation of red giants in both evolutionary stages and theoretical stellar models \citep{mosser_spin_2012,eggenberger_angular_2012,marques_j_p_seismic_2013, Deheuvels2015, Gehan2018, Mosser2024}. Theoretical models including additional transport processes so far predict a spin-up of the core as core-helium burning proceeds \citep{Moyano2022,Moyano2023}, which goes against recent observations by \citet{Mosser2024}.

{In addition, numerous studies have highlighted the need for additional mixing at the border of convective regions, e.g. at the base of the convective zone in H-shell burning red giant stars \citep{lagarde_thermohaline_2012,Khan2018} or at the top of the convective core of main-sequence stars with masses higher than $\approx 1.2 M_{\odot}$ \citep[][amongst others]{Claret2016, deheuvels_measuring_2016,Claret2019,2020ApJ...904...22V,noll_probing_2021,2023A&A...676A..70N,2024ApJ...965..171L}. During core helium burning, a similar situation occurs with the presence of a growing convective core, the evolution of which may be altered by additional mixing at its border \citep{montalban_testing_2013,Bossini2015,vrard_evidence_2022,Dreau2022}.} Despite the importance of core helium burning stars for understanding chemical mixing as their core is the site of convective motions, the complexity of their oscillation spectra has led to a lack of detailed studies on these promising targets. Computing models suitable for detailed asteroseismic applications during this evolutionary phase is challenging, notably due to the presence of sharp variations in the structure of the inner core \citep{vrard_evidence_2022} which create glitch signatures \citep[see ][for analytical developments and theoretical discussions on buoyancy glitches]{cunha_structural_2015,Cunha2019,Cunha2024} in their seismic spectra, to which is added the problem of semi-convection (see e.g. \cite{salaris_chemical_2017} and references therein). Some works have even looked at the seismic properties of stars going through the helium flashes \citep{Deheuvels2018}. Extracting valuable information from asteroseismic data requires advances in understanding which processes generate the observed signals and determining the most relevant seismic constraints to accurately infer the internal structure. Therefore, additional mixing processes such as convective overshooting and semi-convection have yet to be characterised for this evolutionary phase using detailed asteroseismic analyses. \\
In this context, we present theoretical asteroseismic applications of a new version of the Liège stellar evolution code \citep[CLES,][]{Scuflaire2008} for evolutionary models of core-Helium burning stars of different initial parameters where their mode frequencies are computed with the Liège adiabatic oscillation code \citep[LOSC,][]{Scuflaire2008}. The evolution code provides a consistent treatment of semi-convection and a high degree of consistency regarding the verification of hydrostatic equilibrium and the treatment of convective boundaries that allows detailed asteroseimic analyses. Our aim is to establish a framework for detailed asteroseismic analysis of core-helium burning stars, following the idea of providing additional insights to the theoretical models currently used in stellar evolution codes and hydrodynamical simulations \citep[e.g.][amongst other]{Mirouh2012,Cristini2016,Rizzuti2022,Blouin2024}. In particular, we aim to investigate how the asteroseismic properties of these stars are influenced by semi-convective and overshooting processes, focusing on the connection between seismic signatures and internal chemical composition. Notably, our study will examine mode trapping due to sharp structural variations, with a trapping region being a zone where an oscillation mode remains confined, with large amplitudes inside and rapidly decreasing outside, bounded by turning points where reflection occurs \citep[see e.g.][for a discussion of dipolar mixed oscillation modes]{Takata2016I,Takata2016II}.

{Modelling low-mass red giants in the core-He burning phase presents significant challenges due to their complex internal structure. 
These difficulties become particularly evident after the initial stage of core-Helium burning, when the semi-convective layer emerges.
The treatment of this region in stellar evolution computations requires us to manage the \textit{semi-convection problem}. We recall that a region that undergoes convective motions is characterised by the Schwarzschild criterion $\nabla_{ad} < \nabla_{rad}$, which expresses convective instability due to a high temperature gradient through the radiative and adiabatic gradients, defined in Sect.\ref{sect:indicators}. To ensure a coherent model, the boundary of the convective core must satisfy that the resulting force (hence acceleration) must vanish, $\nabla_{ad} = \nabla_{rad}$ on the convective side of the boundary. In CLES, we follow the prescriptions described in  \citet{Gabriel2014}. Beyond the location of core boundary, an additional region affected by convective elements may appear, the so-called overshooting region, which is also a critical issue when modelling main sequence stars exhibiting a convective core. The challenge in modelling a core-He burning star lies notably in handling the growth of the convective core and the appearance of the semi-convective zone \citep[e.g.][]{castellani_helium-burning_1985,straniero_chemical_2003,serenelli_constructing_2005,Noels2013}. As discussed in \cite{salaris_chemical_2017} and analysed in the work of \cite{castellani_induced_1971}, the initial phase of core-helium burning is characterised by a convective core whose mass is growing with time due to the increase in opacity as helium is transformed into carbon and oxygen. The radiative gradient develops a minimum located inside the fully mixed core, which lies at the root of the semi-convection problem. In practical terms, this minimum leads to an ambiguous boundary for the convective core as the radiative gradient increases beyond the point of neutrality. \cite{salaris_chemical_2017} provides a visualisation of the gradients profile in their Section 4.2, Fig. 8 and 9. We also refer the readers to \citet{Castellani1971,Castellani1972,Castellani1982,straniero_chemical_2003,Giammichele2018} for additional references on the evolution of convective cores during core-He burning and to \citet{Girardi2016} for a recent review.}

In Sect.\ref{sect:evolution}, we present our stellar models and details about the formalism used in the stellar evolution code to cover this stage of evolution. We introduce the global asteroseismic parameters in Sect.\ref{sect:indicators} and present the seismic signature of structural internal variations during the evolution, related to mixing processes such as convection and semi-convection. In the last section (Sect.\ref{sect:analysis}), we focus on the analysis of the oscillation spectra of various models to give an overview of the expected asteroseismic signatures of our treatment of semi-convection and overshooting over our considered range of parameters for the models. 

\section{Stellar evolution models} \label{sect:evolution}
 
\subsection{Liège Stellar Evolution Code (CLES)}
In the code CLES \citep{Scuflaire2008}, a semi-convective zone appears above the overshooting region (whenever overshooting of convective elements is considered). The base of the semi-convective region is located beyond the minimum of the radiative gradient at the point where it reaches the adiabatic gradient. By adjusting the chemical composition, this semi-convective region ensures that the neutrality condition $\nabla_{ad} = \nabla_{rad}$ is satisfied from that point to the discontinuity boundary, creating a chemically variable profile. This approach known as \textit{induced semi-convection} is similar to methods used in other stellar evolution codes \citep{Ventura2008,Deglinnocenti2008,Bressan2013} and is the one implemented in our code.

{Since the original publication in 2008 \citep{Scuflaire2008}, the software has seen numerous changes. In this new version, we adopted the same strategy for the
helium burning phase and developed a new branch dedicated to it. We did not include
the description of phenomena that would not alter the conclusions of our work, such as mass loss and diffusion, in order to focus our attention and efforts on
convergence problems and the description of the semi-convective zone. In this
way, we were able to test different prescriptions for the composition and
boundaries of the semi-convective zone.}

{Although the helium combustion reactions were coded in the original version of
CLES, this version was not suitable for calculating these evolutionary phases,
and the calculation had to be stopped as soon as helium combustion begins. The
rest of the evolution is based on two programs, ZAHB and HBevol. If the mass of
the model is sufficient to avoid helium flash, the rest of the evolution is
handled by the HBevol program. For smaller masses,  we start just after the flash and use ZAHB (Zero Age Horizontal Branch) to
calculate an initial static model, which serves as the starting point for HBevol. These initial models were inspired by the properties of ATON evolutionary models \citep{Ventura2008} that were used in the initial development phases.}

{The ZAHB program computes a static stellar model intended to represent the star immediately after the helium flash. This approach is justified by the fact that, during the central helium-burning phase, the gravitational energy contribution is negligible compared to the nuclear term. The model is defined by the usual parameters: total mass, initial hydrogen and metal fractions ($X$, $Z$), and convection and overshooting parameters, together with two core parameters that determine its internal properties: the helium core mass ($m_{He}$) and the carbon enrichment of the helium core XCO produced during the flash. In fully evolutionary computations, these parameters are direct outputs of the stellar evolution, but here they are treated as free parameters defining the initial conditions for the core-helium burning phase.}

{The parameter $m_{He}$ generally lies between 0.45 and 0.50 $\mathrm{M}_\odot$; as a first step, we fix it to 0.50 $\mathrm{M}_\odot$, close to values found in evolutionary models \citep{Girardi2016}. A later step in our work will be to calibrate this value from full evolutionary calculations.} By keeping the mass of the helium core constant while varying the total mass, our approach essentially corresponds to physically altering the envelope mass of the models and computing their evolution right after they have undergone the flash. This approach allows us to reduce the complexity of the analysis and focus on the effects of the other free parameters. The second parameter, XCO, denotes the mass fraction of helium converted to carbon through the triple alpha nuclear reaction, leaving in our modelling approach a small discontinuity in chemical composition, located below the hydrogen-burning shell and attributed to the remnant of the \textit{helium-flash} \citep{kippenhahn_stellar_2013}.
 
\subsection{Models computation}

The stellar models used in this study were computed with the code CLES using the FreeEOS equation of state \citep{Irwin}. Their initial properties are the following. The CHeB models we compute are at solar metallicity, taken from \citet{asplund_chemical_2009}. The mass fraction of hydrogen and metallicity at the surface are fixed at 0.72 and 0.014, respectively. Inside the core, the mass fraction of carbon, determined by the parameter XCO, is set at 0.05 at the beginning of our computations.

The mass of the helium core is a free parameter that we initially fixed at $m_{He}$ = 0.50$\rm{M}_{\odot}$ for all our model, then CLES computes the evolution coherently.  Opacities are taken from type II OPAL tables \citep{iglesias_updated_1996}, supplemented at low temperature by the Wichita State University opacities  \citep{ferguson_low-temperature_2005}. Our models exhibit a convective core surrounded by an overshooting layer, a semi-convective layer modelled using induced-convection, a hydrogen-burning shell and a convective envelope modelled by means of the mixing length theory \citep{weiss_cox_2004} with a solar-calibrated mixing length of $l=1.8$ $H_P$  with $H_P=\frac{-dr}{dln P}$ the pressure scale height. The radial extent of the penetrative convection region \citep{zahn_convective_1991}, which undergoes instantaneous mixing and has the temperature gradient set to the adiabatic one \citep{maeder_stellar_1975}, is initially set as $d = \alpha_{ov}$ $H_P$ using a free parameter in the code denoted $\alpha_{ov}$, the overshooting parameter. At the surface, the models have an Eddington gray atmosphere.

\begin{figure}[]
    \centering
    \includegraphics[width=0.99\linewidth,keepaspectratio]{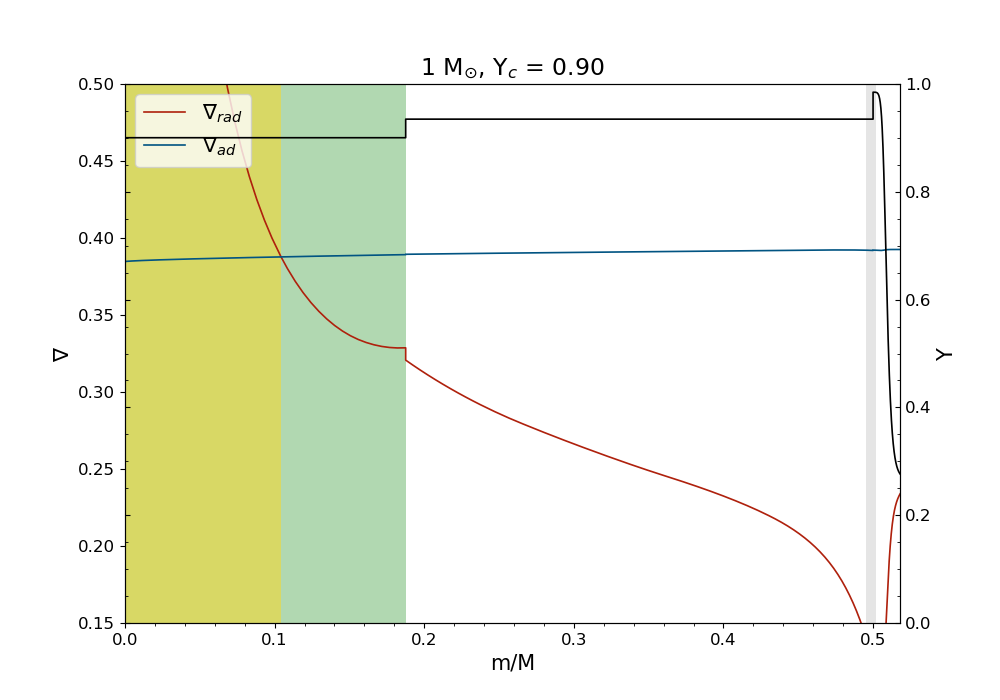}
    \caption{Gradients profile as function of the mass fraction for a one solar mass CHeB star at solar metallicity and Y$_c$ = 0.90. The helium profile is put in parallel (black). The convective zone is in yellow, the overshooting layer in green and the sharp variation left by the \textit{He-flash} in grey with XCO = 5\%.  The radiative gradient $\nabla_{rad}$ is in red and the adiabatic gradient $\nabla_{ad}$ in blue. The real gradient $\nabla$ is defined as following, $\nabla = \nabla_{ad}$ in the \textit{COS} zone and $\nabla = \nabla_{rad}$ in the radiative zone. We focus on the central part of the star for showing purpose.}
    \label{fig:helium-flash}
\end{figure}

The discontinuity in chemical composition due to the depletion of helium (XCO) that we associate with the \textit{helium-flash} creates a sharp transition in the helium profile, as illustrated in Fig.\ref{fig:helium-flash} around $0.5$ m/M. We repeat here that XCO is so far a free parameter in our code but can be calibrated using consistent evolutionary computations. In our case, to keep more flexibility in our investigations, we keep it as a free parameter of the computations. 

In the following sections, we call the \textit{COS} zone the set of the 3 central zones (Convective core, Overshooting, and Semi-convective layers), and the \textit{fully mixed} core the combination of the convective core and the overshooting layer.

\begin{figure} 
\begin{subfigure}{0.49\textwidth}
	\centering
		\includegraphics[width=8.5cm]{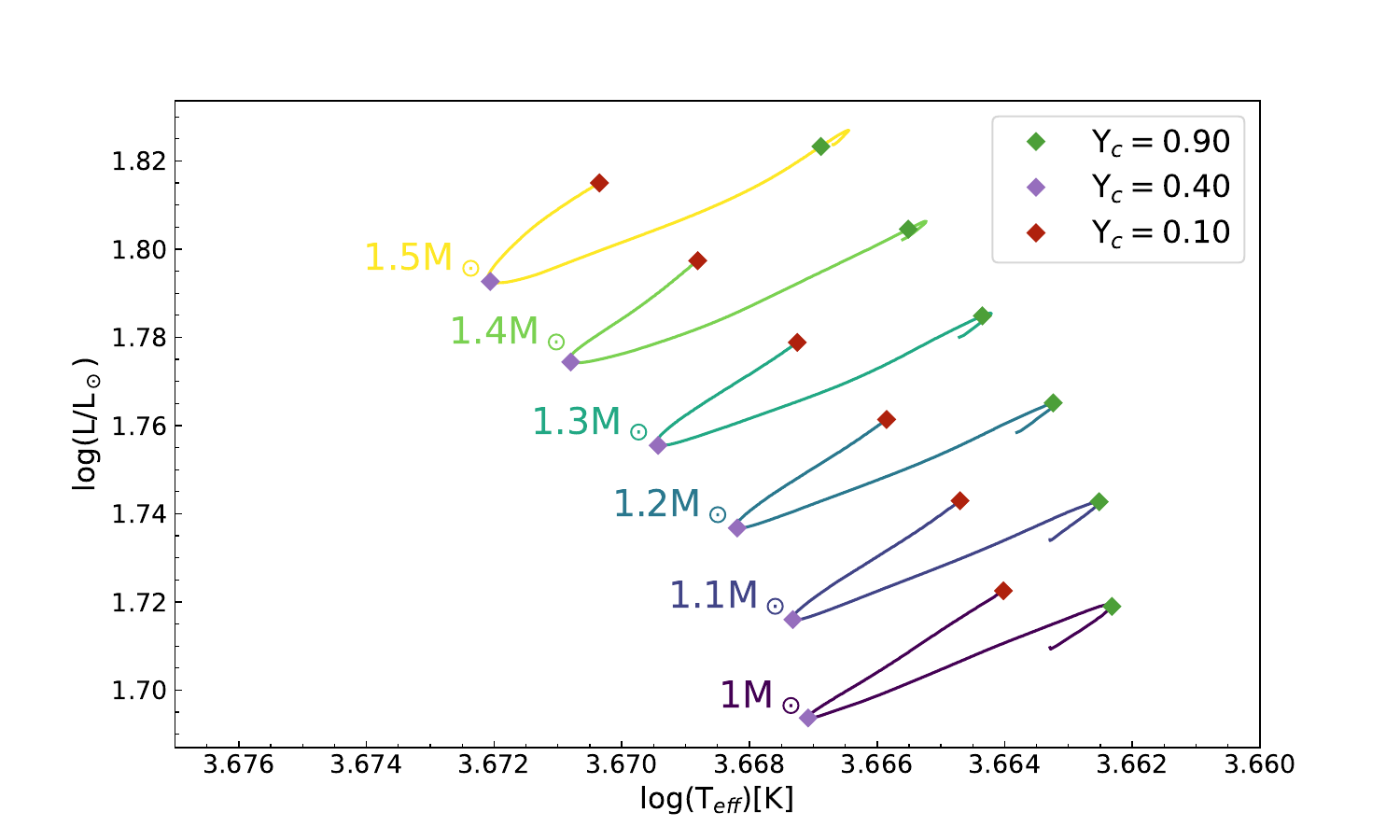}
\end{subfigure}
\begin{subfigure}{0.49\textwidth}
\centering
    \includegraphics[width=8.5cm]{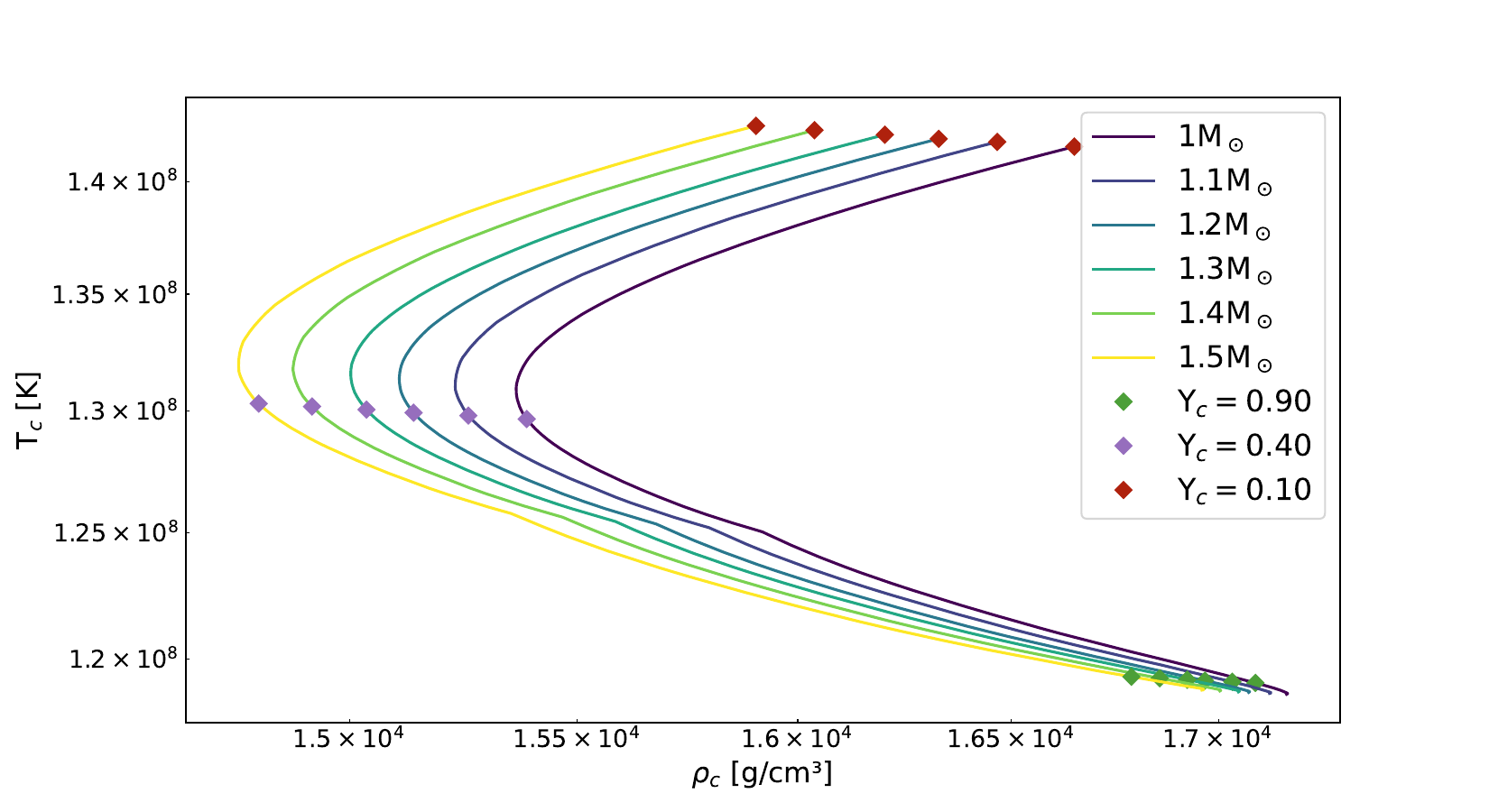}
		
\end{subfigure}
\caption{Top panel: evolutionary track in the Hertzsprung–Russell diagram of the sequence of models from 1 M$_\odot$ to 1.5 M$_\odot$ stellar masses at fixed He-core mass. The markers indicate the state of central helium abundance studied Y$_c$ = 0.90,Y$_c$ = 0.40 and Y$_c$ = 0.10. Bottom panel: central temperature as a function of the central density in logarithmic scale of the corresponding set of masses.}
\label{Fig:hr}
\end{figure}

The evolutionary track of our models in the Hertzsprung–Russell diagram follows a similar pattern to that of evolutionary models of core-helium burning stars for masses from 1 M$_\odot$ to 1.5 M$_\odot$\footnote{At higher masses, our input parameter describing the helium core mass would need to be altered for the models to remain in line with the predictions of evolutionary computations.} \citep[see e.g.][for a review]{Girardi2016} as illustrated in the top panel of Fig.\ref{Fig:hr}. The total luminosity gradually decreases on the horizontal branch, while the effective temperature remains almost constant. At a state around Y$_c= 0.40$, a turn-around occurs in the track and the luminosity starts to increase. This characteristic may be sensitive to the treatment of convective overshooting or other mixing processes. For comparison, it occurs at Y$_c \sim 0.30$ in \citet{montalban_testing_2013}, and around Y$_c \sim 0.75$ in \citet{constantino_treatment_2015}. These differences highlight the need for further investigation, especially given the uncertainties associated with the treatment of the mixed regions close to the helium burning core and the way the Schwarzschild criterion is applied in various evolutionary computations \citep{Gabriel2014}. 
 
During this phase, luminosity comes from two main sources of energy, the H-burning shell surrounding the He-core and the helium burning at the centre. As helium is transformed into carbon in the core, the opacity increases, which results in an increase of the radiative gradient and of the convective core mass. The temperature sensitivity of the 3$\alpha$ reactions is so large that it imposes a decrease in density \citep{kippenhahn2012}. The reactions are challenged by the production of oxygen with the reaction $\leftidx{^{12}}{C}(\alpha,\gamma)\leftidx{^{16}}{O}$ becoming progressively dominant.
 
Simultaneously, the H-shell fades away, causing a decrease of the surface luminosity. When Y$_c$ drops to around 0.40, the central density starts to increase, as shown in the $\rho_c-T_c$ diagram in the bottom panel of Fig. \ref{Fig:hr}. This turn-around is linked to the appearance of the semi-convective layer since it limits the growth of the fully mixed core in our models. This changes the amount of mass taken in the convective core, influencing the nuclear reactions as treated in CLES \citep{angulo_compilation_1999, Kunz2002} and consequently the evolution of the core and the central density.

The core contraction raises the temperature in the shell, increasing the nuclear rate of hydrogen burning and increasing both shell and surface luminosity. As the core approaches the helium abundance of Y$_c = 0.10$ the production of oxygen becomes more and more important as the contribution of  the 3$\alpha$ reaction decreases, causing the core to contract further and the envelope to expand.  

 {The induced semi-convection method implemented in CLES imposes the chemical composition gradient to obtain the neutrality condition of the gradients $\nabla_{ad}=\nabla_{rad}$.This is done by allowing a partial chemical mixing just enough to bring the radiative gradient equal to the adiabatic value and imposing conservation of the total mass of each chemical element (apart from nuclear reactions)} Although this provides a consistent framework for modelling the semi-convective regions in core-helium-burning stars, our understanding and modelling of the underlying physical processes remain incomplete. Various approaches can be considered to address this issue, such as 3D hydrodynamic simulations as in, e.g., \cite{Wood2013}. More recent studies, such as \cite{Blouin2024}, examine the stability of semi-convective layers, while \cite{fuentes_3d_2025} investigate rotation effects in semi-convective transport, although in the case of Jupiter in the latter work. While these simulations provide insights into the dynamics of semi-convection, they remain subject to some limitations. They are not tested against observational constraints or carried out in fully stellar conditions (e.g., thermal relaxation over evolutionary timescales),  and some might not include physical processes such as rotation and magnetism, which could alter the efficiency of the mixing. These studies and their different conclusions regarding semi-convective regions highlight the need for further investigations in stellar interiors. A complementary approach is to use seismic constraints as a tool to extract detailed information on the internal structure and mixing processes within these regions, using the information provided by individual oscillation modes, which is the primary focus of our study.

\section{Global asteroseismic indicators} \label{sect:indicators}

According to asymptotic theory, global seismic indicators can be defined for both the pressure- and gravity-modes, see \citet{Shibahashi1979,Tassoul1980} and the works of \citet{Takata2016I,Takata2016II} for mixed oscillation modes. 

The large frequency separation is expressed as:
\begin{equation}
    \Delta \nu_{n,\ell} = \nu_{n,\ell} - \nu_{n-1,\ell} \simeq \left( 2\int_0^R \frac{\text{d}r}{c}\right)^{-1},
\end{equation}
with $c$ the local sound speed, the integral being the acoustic radius, $r$ and $R$ the local radial coordinate and the stellar radius, respectively. $\ell$ and $n$ are the degree of the spherical harmonics and the radial overtone, respectively. The asymptotic large frequency separation is a constant commonly used in red giants and is defined from radial modes which are pure p-modes and correspond to $\ell=0$ \citep[see][for a discussion]{Mosser2013}. 

Gravity-dominated mixed modes are typically associated with quasi-equidistant periods defined as the period spacing
\begin{equation}\label{int}
\Delta  \Pi_{n,\ell} =  \Pi_{n,\ell} -  \Pi_{n-1,\ell} \simeq \frac{2 \pi^{2}}{\sqrt{\ell(\ell+1)}}\left( \int \frac{N}{r} \text{d}r\right)^{-1},
\end{equation}
with $N$, the Brunt-Väisälä frequency defined as
\begin{align}\label{brunt}
    &N^2 = \frac{-G m}{r^2}\left(\frac{d \ln \rho}{dr} - \frac{1}{\Gamma_1}\frac{d \ln P}{dr}\right)\approx \frac{\rho g^2}{P}(\nabla_{ad} - \nabla + \nabla_\mu).
\end{align}
$g$ is the local gravitational acceleration and $ \nabla,\nabla_{ad}$, $ \nabla_\mu$  respectively are the temperature, adiabatic and mean molecular weight gradients $\nabla = \frac{d\ln T}{d \ln P} \ , \ \nabla_{ad} = \left.\frac{\partial \ln T}{\partial \ln P} \right\vert_S \ , \ \nabla_\mu = \frac{d\ln\mu}{d\ln P}$ . We compute the integral only in regions where $N^2$ is higher than zero. 

The large frequency separation of radial modes ($\ell=0$) {does not vary with the frequency} in the asymptotic regime. From now on, we will consider this constant asymptotic value and note it $\Delta \nu$. Similar behaviour is expected from the period-spacing of dipolar modes in the asymptotic regime, therefore we define an asymptotic period-spacing of dipolar modes ($\ell=1$), called $\Delta \Pi$ that will be used to study the properties of our models. Both quantities are global seismic indicators commonly used when studying red giants as they can often be determined with a high degree of precision \citep[typically of the order of $0.1\mu Hz$ for $\Delta \nu$ and $\approx 3s$ for $\Delta  \Pi$ as in][]{vrard_evidence_2022,Mosser2024Poster}. In our grid of models, these parameters can be analysed throughout the entire evolutionary sequence. Internal structural variations, particularly those affecting the chemical gradient, influence the {gravity-mode cavity} of mixed modes.

\begin{figure}[!h]
\begin{subfigure}{0.49\textwidth}
\includegraphics[width=9.5cm]{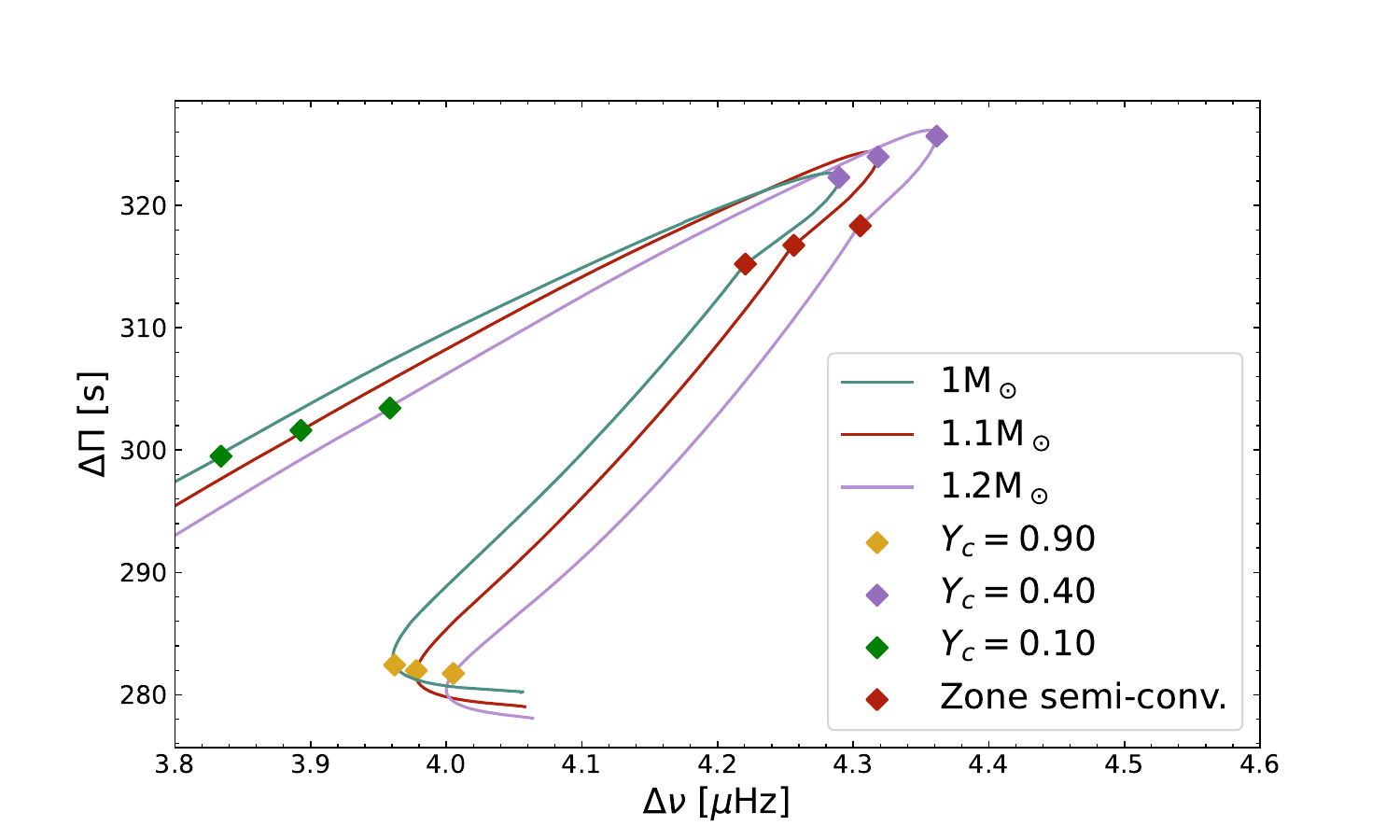}
\caption{Models mass : 1 M$_\odot$, 1.1M$_\odot$, 1.2 M$_\odot$.}\label{fig:mass}
\end{subfigure}
\begin{subfigure}{0.49\textwidth}
\centering
\includegraphics[width=9.5cm]{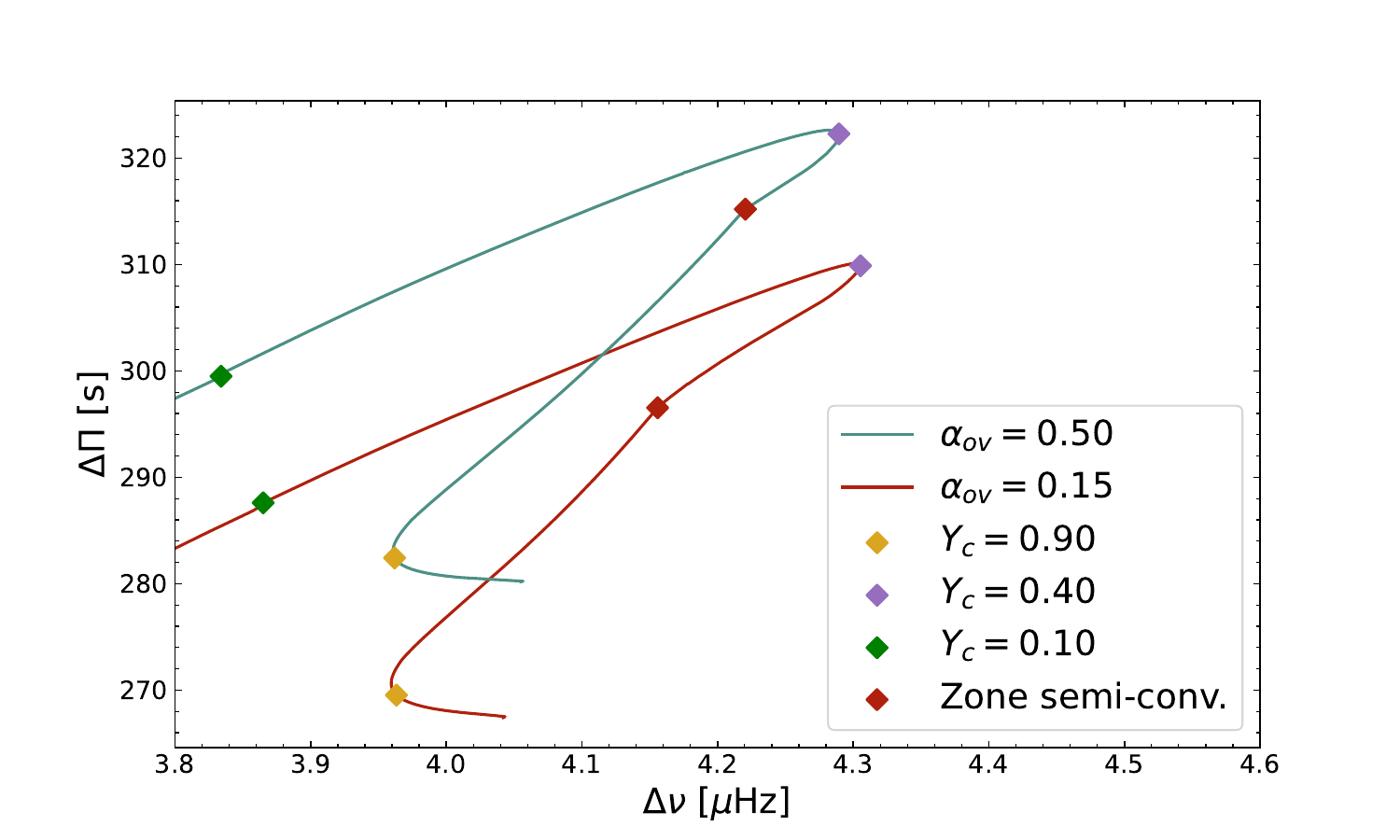}
\caption{Models 1 M$_\odot$ and $\alpha_{ov} = (0.15,0.50)$.}\label{fig:over}
\end{subfigure}
\caption{ Top panel: diagram $\Delta \nu-\Delta  \Pi $ of three models of different stellar mass. Each point correspond to a stage Y$_c$ in the evolution on the horizontal branch. The markers indicate the state of central helium abundance  Y$_c$ = 0.90 ,Y$_c$ = 0.40, Y$_c$ = 0.10 and the appearance of the semi-convective layer. Bottom panel: diagram $\Delta \nu-\Delta  \Pi $ of two models of different overshooting parameter $\alpha_{ov}$.}
\label{Fig:dpir}
\end{figure}

We start by analysing the global asymptotic seismic parameters of a selected grid of models of mass 1 M$_\odot$  with $\alpha_{ov}=0.50$ using the seismic diagram $\Delta \nu - \Delta  \Pi $ shown in Fig. \ref{fig:mass}. Starting from early stages on the horizontal branch, $\sim$ Y$_c$ = 0.90, up to a value around Y$_c$ = 0.40, the asymptotic period-spacing $\Delta  \Pi$ increases significantly, similarly to previous studies using evolutionary models \citep[see e.g.][]{montalban_testing_2013,Bossini2015,constantino_treatment_2015}.  Beyond this state, a turn-around in the track appears where it starts to decrease until the end of the sequence. This is similar to what is observed in the evolutionary tracks in the top panel of Fig. \ref{Fig:hr} when the core contracts. In this case, the dependence to the radius r and the evolution of the Brunt-Vaisälä frequency profile can explain this trend in the period-spacing. As mentioned in Eq.\ref{brunt}, $N$ strongly depends on the factor $\rho g^2/P$ that guides the variation of period-spacing as the core expands or contracts.

\begin{figure}[!h]
\begin{subfigure}{0.49\textwidth}
\includegraphics[width=0.99\linewidth,keepaspectratio]{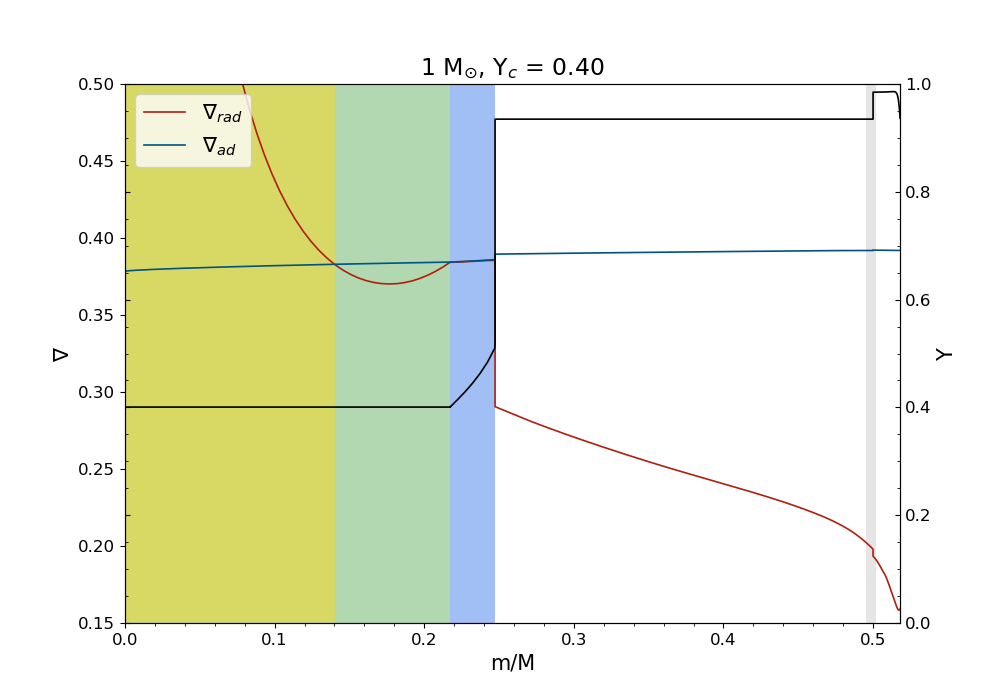}
\caption{Model $\alpha_{ov} = 0.50$ and Y$_c$ = 0.40.}\label{fig:grad}
\end{subfigure}
\begin{subfigure}{0.49\textwidth}
\centering
\includegraphics[width=0.99\linewidth,keepaspectratio]{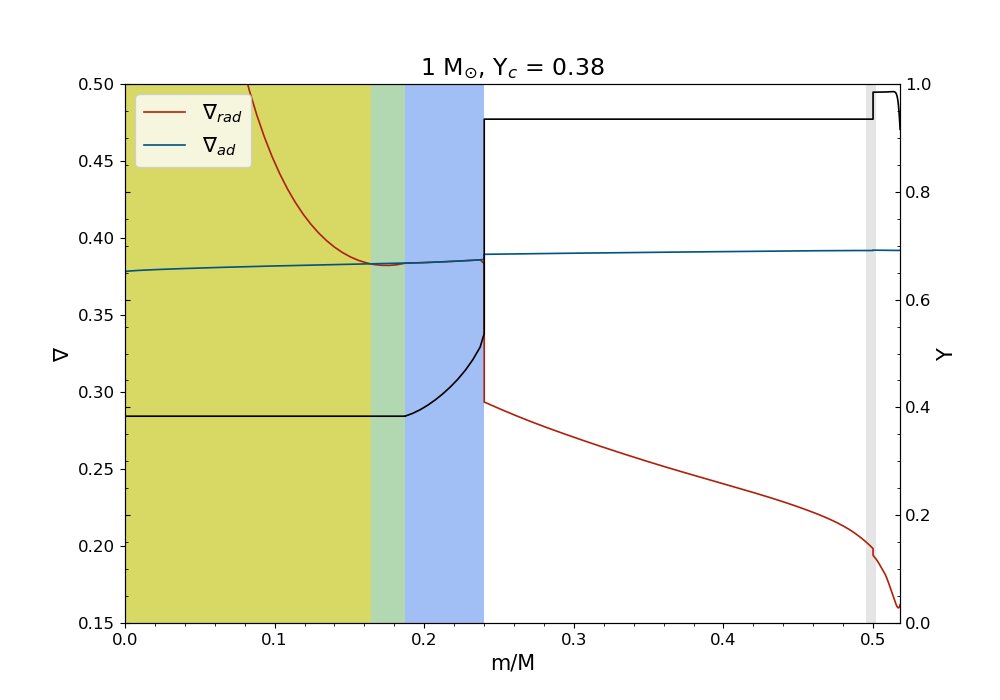}
\caption{Model $\alpha_{ov} = 0.15$ and Y$_c$ = 0.38.}\label{fig:grad-over}
\end{subfigure}
\caption{ Comparison of overshooting parameters $\alpha_{ov} = 0.50$ (top panel) and $\alpha_{ov} = 0.15$ (bottom panel) for 1 M$_\odot$ through the gradients and helium profiles as a function of the mass fraction at same age, in parallel to the helium profile (black). The convective zone is in yellow, the overshooting layer in green, the semi-convective zone in blue and the sharp variation left by the \textit{He-flash} in grey.  The radiative gradient $\nabla_{rad}$ is in red and the adiabatic gradient $\nabla_{ad}$ in blue. The real gradient $\nabla$ is defined as following, $\nabla = \nabla_{ad}$ in the \textit{COS} zone and $\nabla = \nabla_{rad}$ in the radiative zone.}
\label{Fig:GRAD}
\end{figure}

In the \textit{COS} zone, $\nabla$ is set equal to the adiabatic gradient ($\nabla_{ad}=\nabla$) (see Fig.\ref{fig:grad} for the gradients profile) and the chemical composition gradient vanishes {in the fully mixed} region due to efficient mixing considered as a result of convection ($\nabla_{\mu}=0$). Consequently, the Brunt-Vaisälä frequency also vanishes in the fully mixed core and maintains a non-zero profile in the semi-convective zone and radiative parts of the star. As the evolution of the models progresses, around the abundance Y$_c$= 0.40, the core slows down its expansion and, at some point, contracts rapidly, as discussed above when discussing the increasing central density  in the lower panel of Fig.\ref{Fig:hr}. This trend change affects the asymptotic period-spacing as the factor $\rho g^2/P$ increases and the radius $r$ decreases. It draws a turn-around in the track as shown in Fig. \ref{Fig:dpir}, around Y$_c$= 0.40. 
 
\begin{figure}[h]
\centering
    \includegraphics[width=9.5cm]{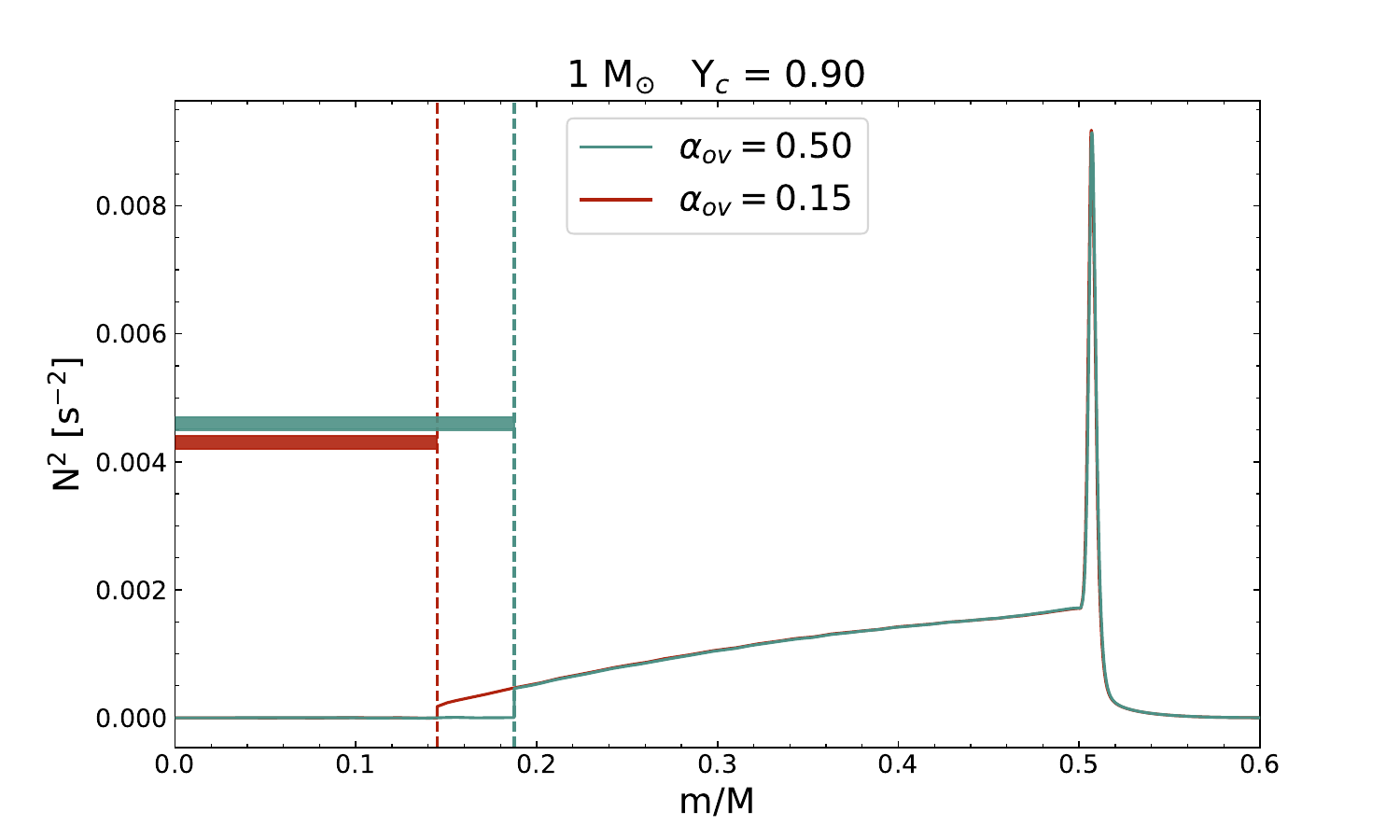}
\caption{Brunt-Väisälä frequency profile of two early models (Y$_c$ = 0.90) for 1 M$_\odot$ with different value of overshooting parameter, $\alpha_{ov}$ = 0.50 (blue) and  $\alpha_{ov}$ = 0.15 (red). The vertical line indicates the fully mixed core boundary of each model.}
\label{Fig:N2}
\end{figure}

Our next investigation is based on our set of models of various stellar masses, in order to analyse the impact on seismic indicators. The study of a slightly higher mass reveals the same general trend of period spacing as the one discussed for 1 M$_\odot$. In Fig.\ref{fig:mass} we select 1.1 and 1.2 M$_\odot$ as comparative example and observe that the tracks have roughly the same initial $\Delta  \Pi$ at same $\Delta \nu$ due to the similar size of the helium cores. The central propagation cavities therefore have the same structure at this particular stage, resulting in a similar profile of $N$ and $\Delta \nu$. We also notice that during the evolution, the tracks depart from each other while the asymptotic period-spacing increases. For the same value of $\Delta \nu$, a lower mass star has a higher value of $\Delta  \Pi$ as the value of $\rho_c-T_c$ is different for each mass. The tracks undergo a change of direction at the same central helium abundance $\sim$ Y$_c$= 0.40 for each model. This corresponds to the same effect as discussed with the evolutionary tracks in Fig.\ref{Fig:hr} with the change of the temperature sensitivity of the central nuclear reactions increasing the central density. 

Another stellar parameter that significantly impacts the global seismic indicators is the overshooting layer. It extends from the convective core where $\nabla_{rad} > \nabla_{ad}$, and variations in its relative size $\alpha_{ov}$, also affect the size of the semi-convective layer in the computed models. Interestingly, the semi-convective layer appears to compensate for variations in $\alpha_{ov}$, such that a larger overshooting layer is accompanied by a smaller semi-convective region and conversely. Fig.\ref{Fig:GRAD} illustrates this trend in parallel to the chemical composition profile. The Y$_c$ when the semi-convective layer appears is also affected with a later appearance for a large overshooting. For models of the same age across the sequence, their helium central abundance is very close. We observe that the total size of the mixed core remains relatively similar once the semi-convective parameter is present, even if the initial extent of overshooting is modified.

The profile of the Brunt-Väisälä frequency undergoes sharp transitions at the bottom and top of the semi-convective region (blue zone in Fig.\ref{Fig:GRAD})  while it is zero in the convective core and the overshooting region where $\nabla=\nabla_{ad}$. As a consequence, variations in the relative size of these regions will affect the seismic parameters. 

We study the asymptotic period spacing $\Delta  \Pi$ by varying the overshooting parameter as illustrated in Fig. \ref{fig:over}. The tracks are shifted, but they are similar in shape. This comes from the variation of overshooting that modifies the initial size of the fully mixed core. In the case of a low overshooting parameter ($\alpha_{ov}$ = 0.15), the  region where $N$ has a non-zero profile is more extended and closer to the centre than the other model with high overshooting ($\alpha_{ov}$ = 0.50). As a consequence, the asymptotic period-spacing $\Delta  \Pi$, is smaller for a low overshooting parameter and this impacts the seismic diagram $\Delta\nu-\Delta  \Pi$. As mentioned above, the semi-convective layer is also affected by this change. A small overshooting implies that the semi-convective region is larger. 

The Brunt-Väisälä frequency profile of early models with Y$_c$ = 0.90 (Fig.\ref{Fig:N2}) shows that the difference of overshooting changes the initial size of the fully mixed core. A model with a lower $\alpha_{ov}$ has a smaller fully mixed core. Therefore, the integral of $N$ in the semi-convective layer is higher, explaining the lower asymptotic period-spacing (Eq.\ref{int}). A similar shift is also observable in the associated seismic spectrum as illustrated in Appendix \ref{A}, Fig.\ref{Fig:spectra_over}. Note that this compensatory effect of the overshooting and semi-convective regions reduces the impact of overshooting on the seismic tracks but even more on the evolutionary tracks in the H-R diagram, where they are almost perfectly superposed. Consequently, while overshooting does have an evolutionary effect, it remains considerably smaller than the impact it has during the main-sequence phase in our models. The seismic effect and signature, on the other hand, depend on the gradient in the overshooting zone emphasising the potential of seismic data to constrain its modelling. 

\begin{figure}[!h]
\begin{subfigure}{0.49\textwidth}
\includegraphics[width=9.5cm]{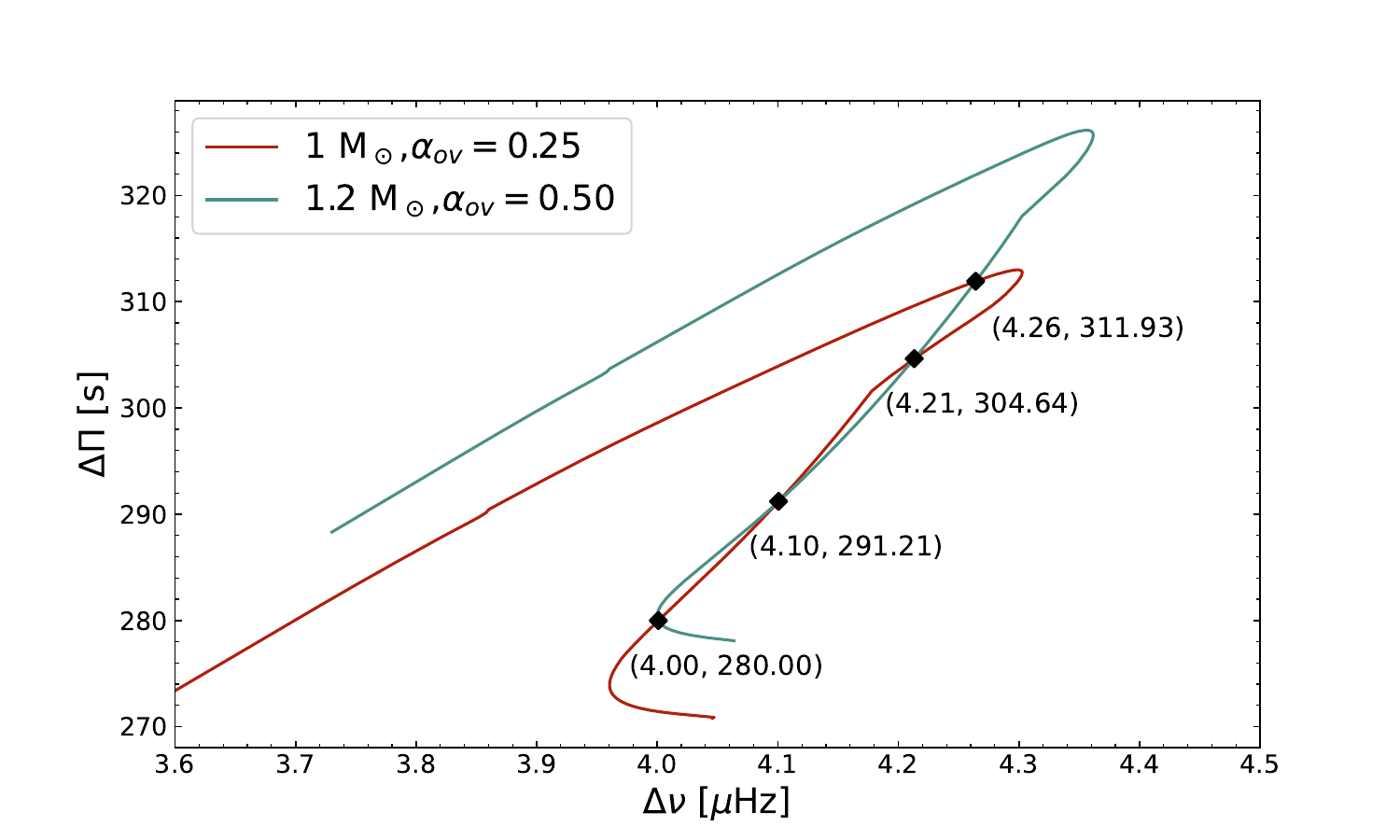}

\end{subfigure}
\begin{subfigure}{0.49\textwidth}
\centering
\includegraphics[width=9.5cm]{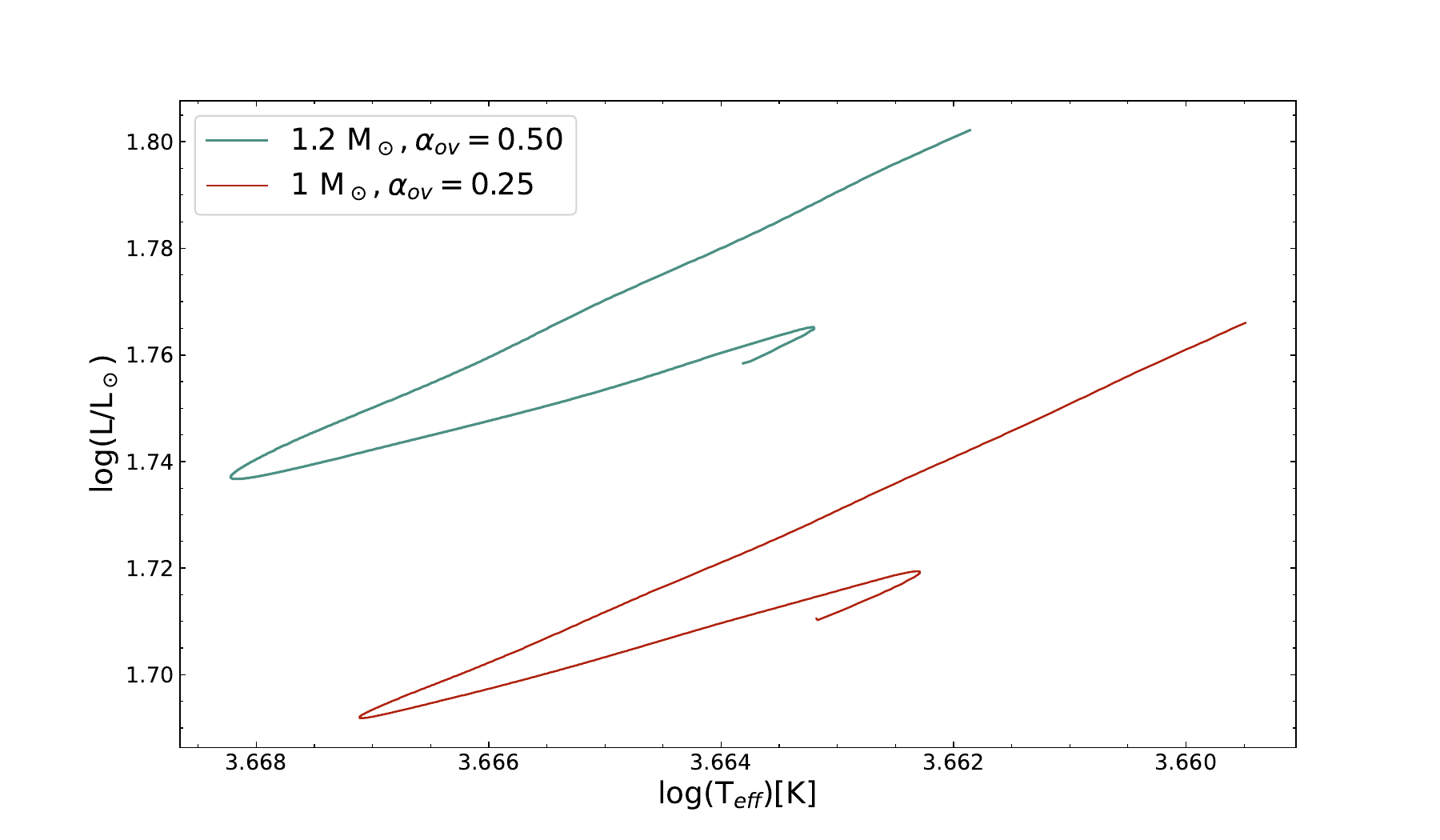}

\end{subfigure}
\caption{ Top panel: diagram $\Delta \nu-\Delta  \Pi $ asymptotic of the models 1.2 M$_\odot$, $\alpha_{ov}$ = 0.50 and 1 M$_\odot$, $\alpha_{ov}$ = 0.25. Each marker corresponds to a crossing of the tracks. Bottom panel: H-R diagram of the corresponding models.}
\label{Fig:degeneracy}
\end{figure}

The result with respect to the seismic parameters raises the question of a potential degeneracy between mass and overshooting \citep[see][for a detailed discussion of between period spacing and core overshooting during core helium burning]{montalban_testing_2013}. Specifically, it suggests that the signature of the mass in the asymptotic period-spacing can be replicated at fixed He-core mass by altering the extent of overshooting. {We computed a model of 1M$_\odot$ with $\alpha_{ov}=0.25$, XCO = 5\% and 1.2M$_\odot$ with $\alpha_{ov}=0.50$, XCO = 5\% .} They show this degeneracy, where the track for one mass intersects that of another when the size of the overshooting region is adapted. For instance, Fig.\ref{Fig:degeneracy} top panel illustrates that the track of the 1.2 M$_\odot$ model with $\alpha_{ov}=0.50$ crosses several times the track of the 1 M$_\odot$ model with $\alpha_{ov}=0.25$ while their H-R diagram tracks are clearly distinct (bottom panel). This observation highlights the need for a more detailed analysis of such stars, as the $\Delta \Pi$-$\Delta\nu$ diagram alone does not reliably distinguish between mass and different mixing prescriptions due to this degeneracy. Given the valuable hidden information carried out on transport processes such as convective overshooting, semi-convective and potential rotational mixing at the border of the helium core, this requires further investigation.

\section{Detailed analysis of the oscillation spectrum} \label{sect:analysis}
 In the previous section, we studied the potential impact of the internal structure and dynamics of a star on the global seismic parameters. In the following section, the oscillation spectra of our model grids are analysed for a specific stage of evolution on the horizontal branch to better grasp the impact of structural changes on the properties of individual oscillation modes. We start with characterisation of their energy density to identify the effect on seismic constraints of the mixing zones near the core. 

Using the eigenfunctions $\xi_r$ and $\xi_h$ describing respectively the radial and horizontal components of the displacement vector $\delta\mathbf{r}(r,\theta,\phi,t)$, we define the normalised inertia of a mode following \cite{christensen-dalsgaard_physics_2004}, as 
\begin{equation} \label{eq:E}
    E = \frac{\int_V \rho|\delta\mathbf{r}|^{2} \text{d}V }{|\delta\mathbf{r}|_{ph}^{2}M}=\frac{\int_M \left(\xi_r^{2} + l(l+1)\xi_h^{2}\right)  \text{d}m}{\left(\xi_{r_{ph}}^{2} + l(l+1)\xi_{h_{ph}}^{2}\right)  M},
\end{equation}
with the integration over the volume of the star and $|\delta\mathbf{r}|^{2}_{ph}$ is the displacement squared norm at the photosphere.  

This approach allows us to identify the trapping properties of the modes or in which part of the star the modes are confined, through the variations in the energy density profile of the selected model and their seismic signatures in the spectrum. It is thus helpful to characterise the relative regions contributing to the mode trapping by extracting and analysing the specific modes that have higher energies than others. This constitutes the first step in identifying the relevant seismic features.
\subsection{He-flash discontinuity} 

\begin{figure}[ht]
    \includegraphics[width=1\linewidth]{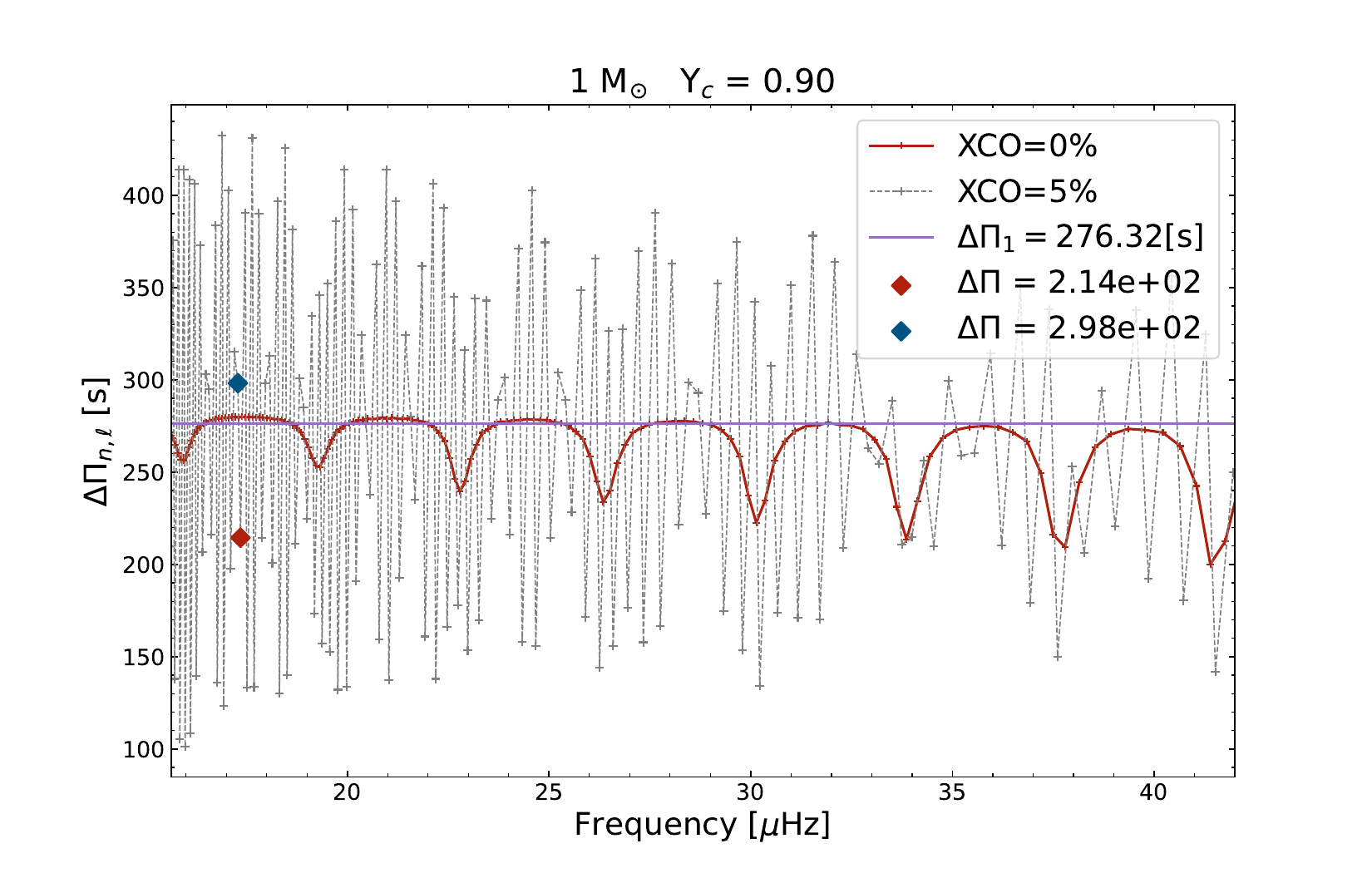}
    \caption{Period-spacing of computed $\ell = 1$ mode frequencies, for models 1 M$_\odot$, Y$_c$ = 0.90 with two different productions of carbon due to the \textit{He-flash}, XCO = 5\% (grey dots) and  XCO = 0\% (red dots), compared with the asymptotic value of the model XCO = 0\% (solid violet line).}
    \label{Fig:spectra-flash}
\end{figure}

Core He-burning stars exhibit variations in dynamics and structure depending on their evolutionary stage and the internal mixing process acting in their core regions. In a solar mass, CHeB star with an early-stage of Y$_c$ = 0.90, the internal structure is relatively simple since the semi-convective layer has not yet appeared. We start by examining the individual period-spacing $\Delta \Pi_{n,l}$ of the computed $l$ = 1 mode frequencies as a function of the frequency $\nu$. A model without any He-flash signature (i.e XCO=0\%), shows similar results to that of a typical shell hydrogen burning red giant (Fig.\ref{Fig:spectra-flash} in red), where local minima correspond to modes trapped in the envelope, and local maxima to modes trapped in the near convective core regions, also called gravity-dominated mixed modes \citep{mosser_period_2015}. Since the chemical gradient $\nabla_{\mu}$ is responsible for mode trapping, it significantly impacts the seismic parameters (see Sect.\ref{sect:evolution} for a discussion) and the variation of overshooting in this model shows similar results to the global period spacing (see Sect.\ref{sect:indicators}) where the spectrum of models with a larger overshooting zone is shifted toward higher asymptotic $\Delta  \Pi$ values. 

As introduced in the previous sections, the \textit{He-flash} discontinuity at the end of the helium core corresponds to a sharp variation of the chemical composition due to the depletion of helium into carbon at the beginning of the core helium burning phase, which also draws a sharp transition in the profile of $N$. By adding the sharp transition due to the \textit{\textit{He-flash}}, where the parameter is denoted XCO=5\%, we observe a significant seismic impact and the expected location of this discontinuity in terms of normalised buoyancy radius \citep[e.g.][]{Miglio2008,cunha_structural_2015,Bossini2015} is perfectly compatible with the periodicity of the glitch signature in the period-spacing. The variation of the intensity of this small discontinuity shows its own signature and seems to be responsible for an additional trapping cavity for the modes as shown in the grey spectrum in Fig.\ref{Fig:spectra-flash} compared to the previous spectrum XCO = 0\% in red. Therefore, the model parameter associated with the \textit{He-flash} or any other mixing responsible for this sharp variation seems relevant to decompose the overall energy density profile.

The corresponding energy density profile associated with{ Fig.\ref{Fig:spectra-flash}} displays two main characteristics. First, it shows a general pattern of local maxima and minima, which is typical of mixed modes and reminiscent of typical shell hydrogen burning red giants, where modes trapped in the deep g-mode cavity with higher inertia and those trapped in the envelope p-mode cavity with lower inertia. Second, in the case of XCO = 5\% the profile features sawtooth patterns overlaid on the main trend. The structure becomes even more complex at low frequencies, featuring a more intricate transition signature.

\begin{figure}[ht]
    \centering
    \includegraphics[width=9cm]{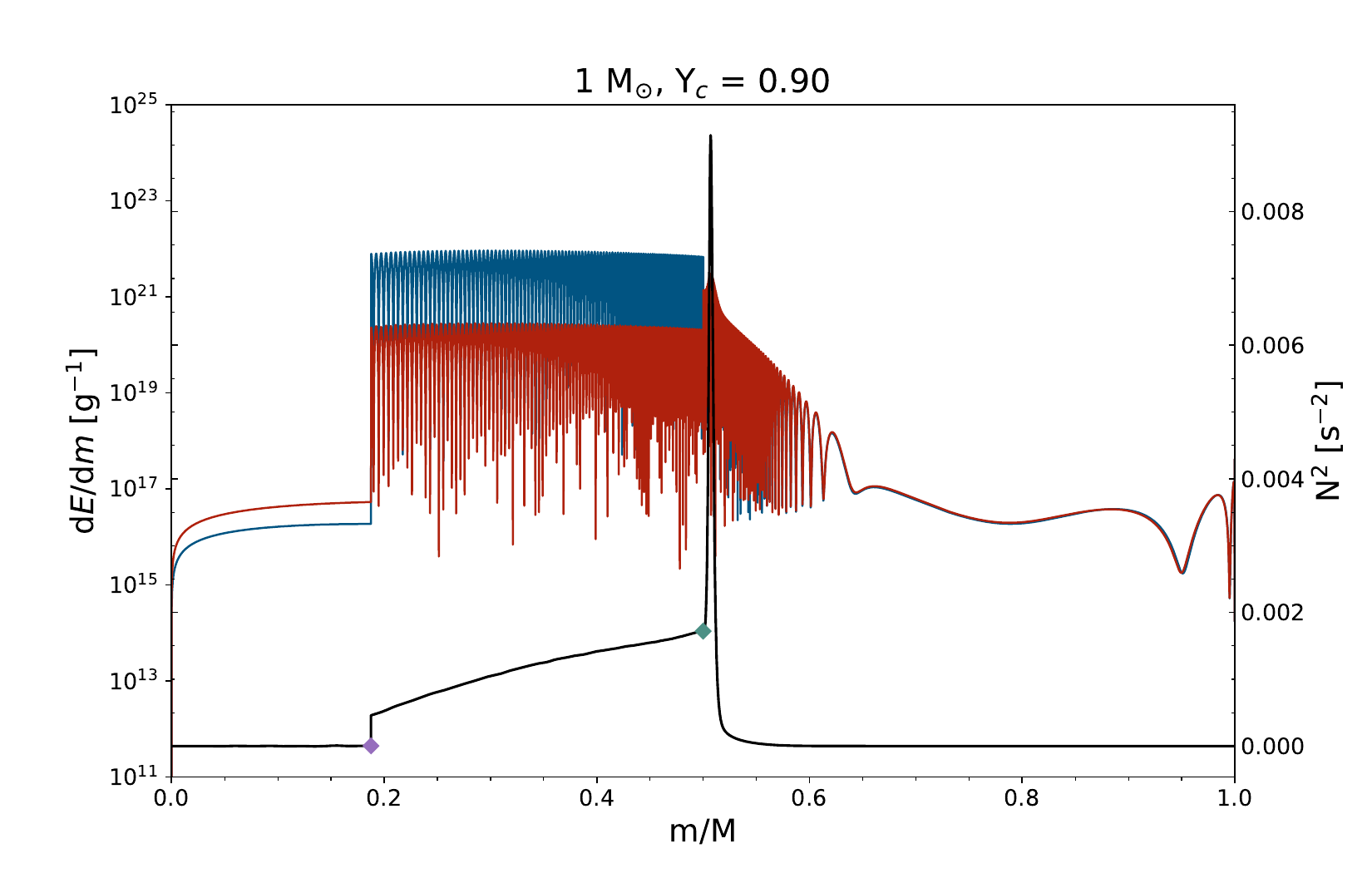}
    \caption{Integrand of energy density in logarithmic scale of two successive modes of the model 1 M$_\odot$, Y$_c$ = 0.90, with the \textit{He-flash} discontinuity and compared to the Brunt-Väisälä frequency profile (black line). Purple and green markers indicate respectively the sharp transitions of the Brunt-Väisälä frequency profile in the core and the \textit{He-flash} discontinuity.}\label{fig:eigen-90}
\end{figure}
To analyse the origin of these sawtooth, we examine the integrand of the energy density, defined in Eq.\ref{eq:E}, for two successive sawtooth modes. The resulting profiles clearly indicate that the mode with a local maximum in energy is trapped within the intermediate radiative zone, which is characterised by what we have called the \textit{He-flash} discontinuity. We show an example of trapping in this specific zone for our model 1 M$_\odot$ and Y$_c$ =  0.90 in Fig.\ref{fig:eigen-90}, in parallel to the corresponding Brunt-Väisälä frequency profile. 
This figure presents two successive modes selected from the same $1,\mathrm{M}_\odot$ model with Y$_c = 0.90$ and XCO = 5\%. These modes, also marked in blue and red in Fig. \ref{Fig:spectra-flash} are indicated in the energy density profile of the same model (Fig.\ref{Fig:inertia-90}) to  illustrate that mode trapping is essentially a question of variations in mode inertia.

The first marker on the Brunt-Väisälä frequency profile corresponds to the sharp boundary of the convective core, where the Brunt-Väisälä profile jumps due to the change to radiative stratification in the model. The second one is linked to the \textit{He-flash} and hides a Dirac delta distribution in the Brunt-Väisälä frequency profile.  The treatment of this discontinuity might not be fully representative of the actual signature of the evolution through the \textit{He-flash}, or the associated various subflashes, but still remain indicative of the impact on seismic constraints of the presence of sharp chemical gradients between the convective helium core and the hydrogen shell.

\subsection{Transition sharpness}
\begin{figure*}[!ht]
\centering
  \includegraphics[width=16cm]{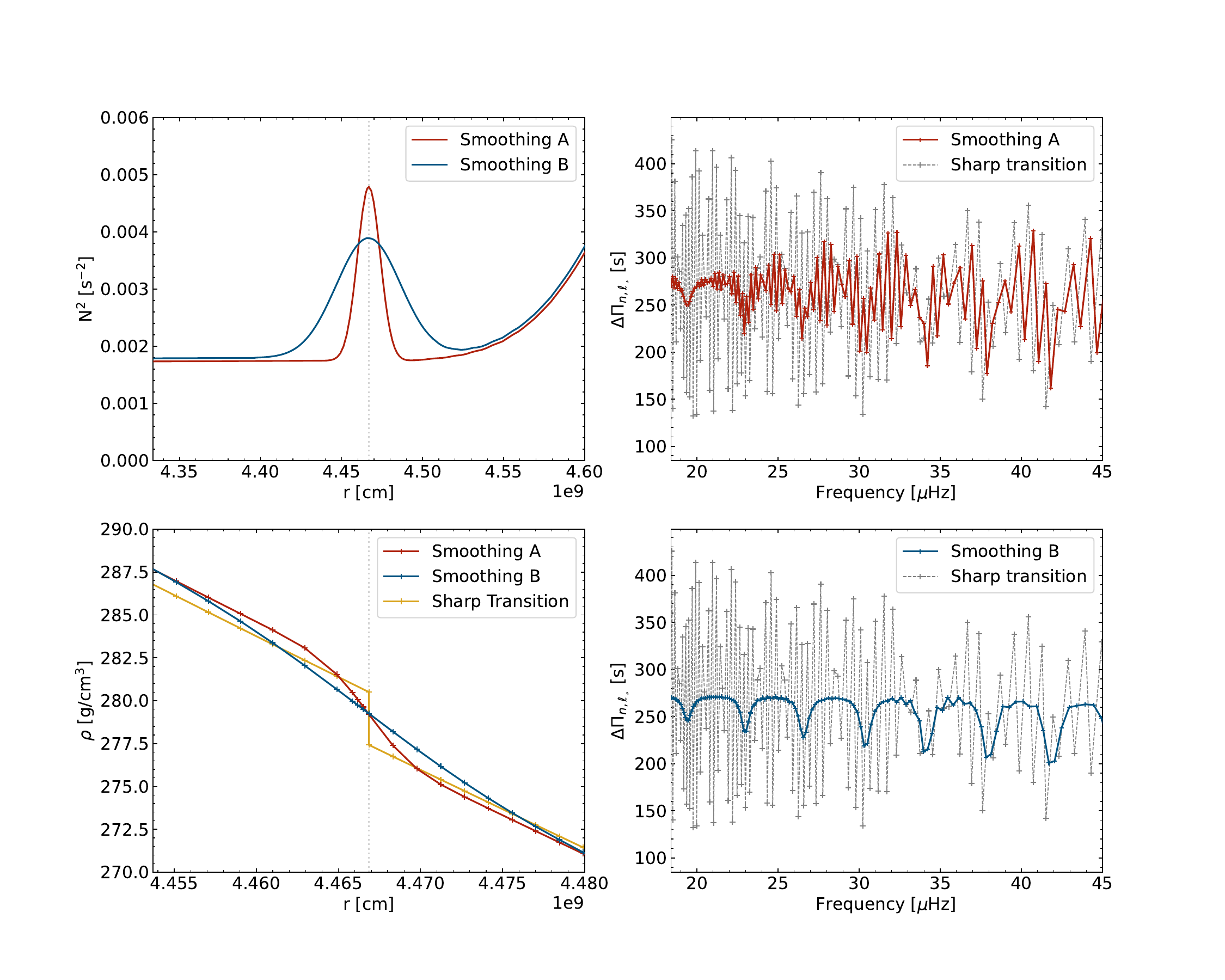}
  \caption{Left-hand panels: density and Brunt-Väisälä frequency profiles of two smoothed models for 1 M$_\odot$, A (Gaussian parameters: amplitude = 9e-10 s$^{-2}$, width = 7e6 cm) and B (Gaussian parameters: amplitude = 6e-10 s$^{-2}$, width = 2e7 cm) around the \textit{He-flash} discontinuity, in parallel to the model studied with a sharp transition. Right-hand panels: period-spacing structure of mode frequencies, for the smoothed models A and B (red dots), in parallel to the model with sharp transition (grey dots).}
  \label{Fig:smoothing}
\end{figure*}

The sharp discontinuity included in CLES at one of the important transition zones in these stars, namely the \textit{\textit{helium flash} }discontinuity, is a prescription applied to simplify the initial conditions of this phase. However, as briefly mentioned above, this sharpness might not fully reflect the actual physical processes that occur in these stars. For example, additional mixing of various physical origins (convective penetration, plumes or rotational mixing) and forms (diffusive or convective) could be responsible for such signatures \citep{2015A&A...580A..61V} but could also smooth out or even erase these discontinuities potentially generated by \textit{sub-flashes} over time, leading to more gradual transitions. 

The question of how a smoother transition impacts the seismic properties of the star can be explored further, and to obtain a first look at how it could influence our seismic study, we implemented a post-treatment to our models to smooth out this transition. The discontinuous helium profile leads to a Brunt-Väisälä frequency exhibiting a Dirac delta distribution at this location as the Brunt-Väisälä frequency is directly linked to the density derivative (see Eq.\ref{brunt}). If we model the glitch in $N$ using a Gaussian approximation, modifying the width and amplitude of the distribution thus changes the transition sharpness in the density profile. We do this within a defined radial interval $\Delta r$ around the transition, where we apply the function $ f(\Delta r) = a \exp\left(-\frac{(\Delta r - b)^2}{2c^2}\right)$, with $a$, $b$ and $c$, respectively, the amplitude, position of the peak and $c^2$ the variance.
We applied this approach to the Brunt-Väisälä frequency (Eq.\ref{brunt}) and recalculated the other structural functions of interest (e.g. density, pressure, $\Gamma_1$) to obtain the theoretical spectrum of our smoothed models. We explored various width values for the parameter $c$, corresponding to different degrees of smoothing, ensuring that the density difference, denoted $\Delta \rho$, in the region immediately before and after the transition remained constant by adjusting the amplitude $a$ of the Gaussian function, accordingly.

Fig.\ref{Fig:smoothing} illustrates two examples of density sharp transition smoothing (smoothing A and B), derived from the corresponding Gaussian approximations of the Brunt–Väisälä frequency, alongside the resulting oscillation spectra before and after smoothing. As anticipated, a lower peak value in $N^2$ leads to a smoother density transition. When extended to larger widths and lower amplitudes, the boundary is simply erased, eliminating mode trapping in this region \citep[see][for details about Gaussian structural glitches]{cunha_analytical_2019}. This explains why the spectrum reduces the amplitude of the additional peaks and reverts to the structure observed in Fig. \ref{Fig:spectra-flash} for the model XCO = 0\%. In contrast, narrowing the peak to an extreme recreates a sharp boundary as the Gaussian converges to a Dirac delta distribution \citep{cunha_analytical_2019}. Another modification consists in altering the radial location of the transition to study its impact on the energy density profile and the modes spectrum. We modified our model by radially shifting the location of the transition, thus reducing the size of the trapping zone between the boundary of the \textit{COS} zone and the \textit{He-flash} transition. This results in a decrease of the number of modes trapped inside this intermediate region. At the level of the precision of \textit{Kepler} data, we expect to be able to disentangle between these various cases with high significance \citep[see][for a recent illustration]{Mosser2024Poster}. Finally, we also tested different resolutions to define the Gaussian function added to the Brunt-Väisälä profile to ensure that the main glitch signatures were not due to numerical issues. 

These comparisons show the sensitivity of asteroseismic constraints to the physical nature of the mixing processes acting in these regions, which may help calibrate in the future the prescriptions used in stellar evolution codes. As we demonstrated, a modification of the chemical gradient resulting from mixing processes, regardless of their origin, directly impacts the observed seismic signature displayed in the models, showing the diagnostic potential of detailed asteroseismic modelling of these stars.

\subsection{Semi-convective zone}

\begin{figure}[h]
\centering
\includegraphics[width=9cm]{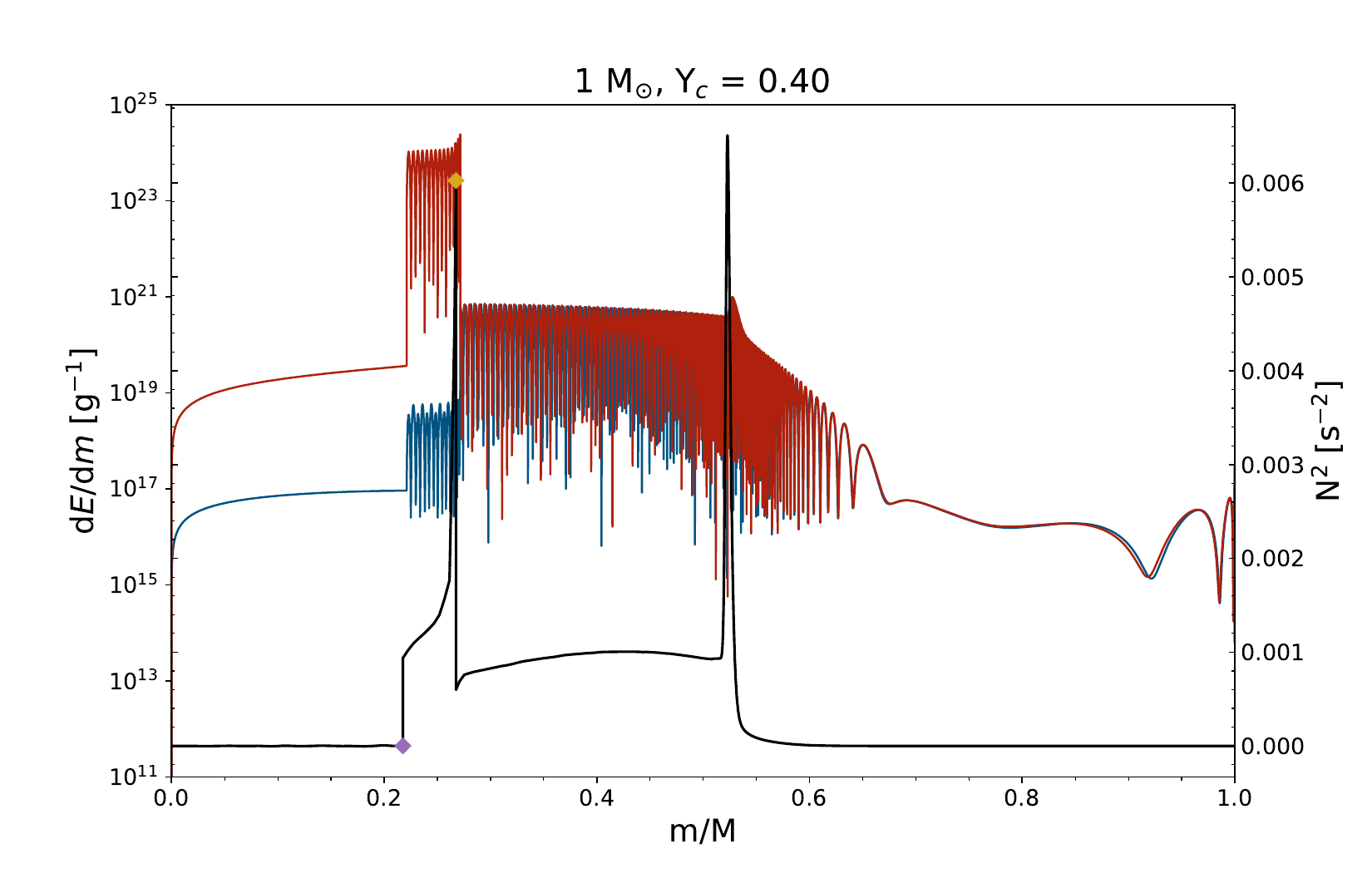}
\caption{Integrand of energy density in logarithmic scale of two successive modes for the model 1 M$_\odot$, Y$_c$ = 0.40, without the \textit{He-flash} discontinuity XCO = 0\% and compared with the Brunt-Väisälä frequency profile (black line). Purple and yellow markers indicate the lower and upper semi-convective boundaries.}\label{fig:eigen}
\end{figure}

\begin{figure}
\centering
    \includegraphics[width=9.5cm]{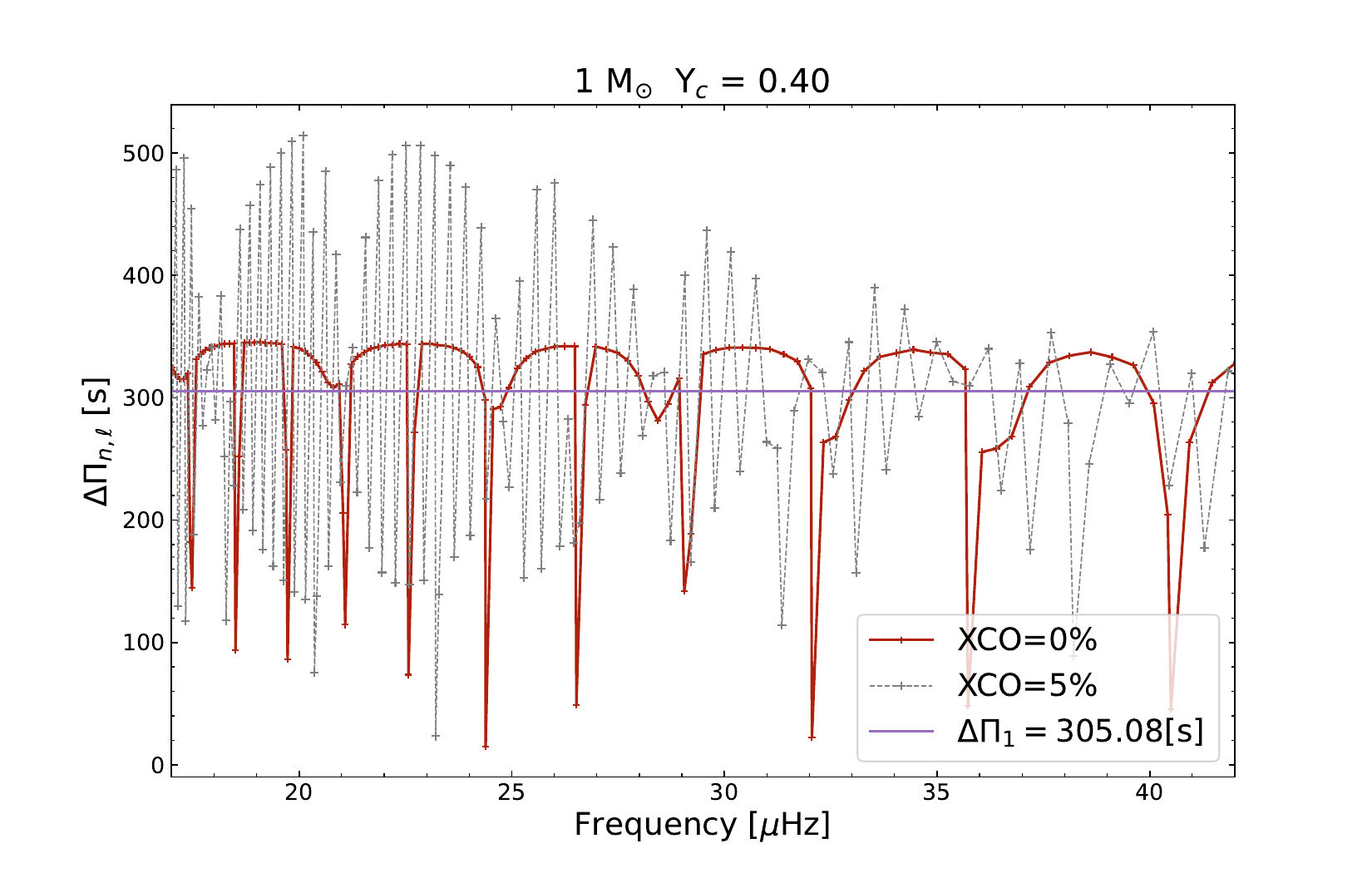}
\caption{Period-spacing of computed $\ell = 1$ mode frequencies, for models 1 M$_\odot$, Y$_c$ = 0.40 with two different productions of carbon due to the \textit{He-flash}, XCO = 5\% (grey dots) and  XCO = 0\% (red dots), compared with the asymptotic value of the model XCO = 0\% (solid violet line).}\label{fig:spectra}
\label{Fig:spectra-sem}
\end{figure}

\begin{figure}[ht]
\centering
    \includegraphics[width=9.5cm]{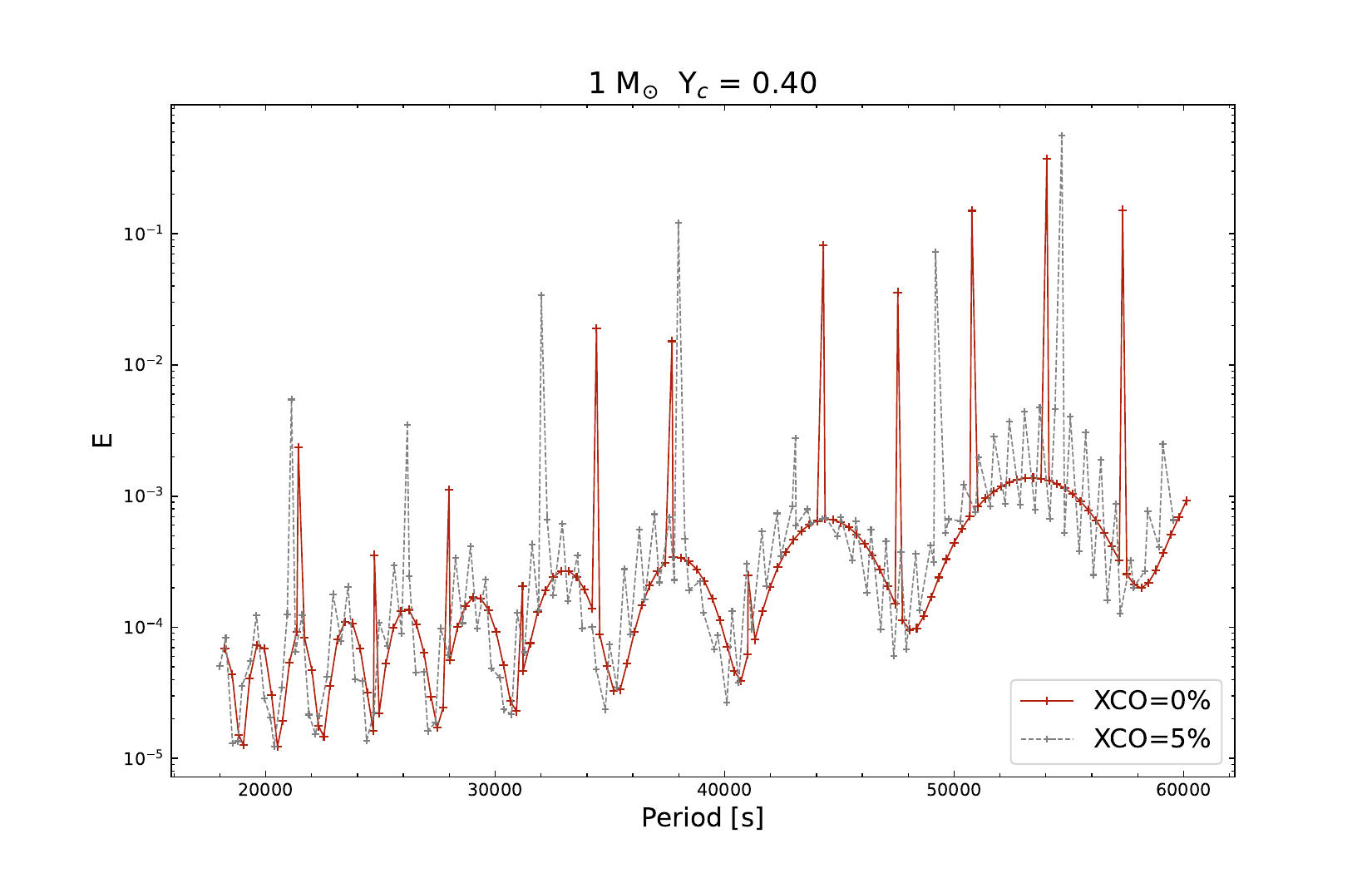}
\caption{Energy density of the modes of models 1 M$_\odot$, Y$_c$ = 0.40 for two sizes of the \textit{He-flash} discontinuity corresponding to fraction of carbon production of XCO = 5\% (grey dots) and XCO = 0\% (red dots).}
\label{Fig:inertia}
\end{figure}

In our grid of models of overshooting $\alpha_{ov} = 0.50$, the semi-convective layer emerges in an advanced evolutionary phase, around a central helium abundance of Y$_c\sim0.56$. This is influenced by the convective overshooting implemented, where a different $\alpha_{ov}$ changes {the} Y$_c$ of the appearance of the semi-convective layer.
The semi-convective layer, region of partial mixing, affects both the chemical gradient and the Brunt-Väisälä frequency profiles, introducing distinct variations in the energy density profile. These signatures appear as sharp energy peaks detached from the main trend, resulting in a more complex oscillation spectrum, even in models where the \textit{He-flash} discontinuity has been smoothed or removed. A detailed analysis of the energy density (Eq.\ref{eq:E}) of the modes associated with these peaks reveals that they are confined within the semi-convective region, delimited by sharp variations in the Brunt-Väisälä frequency profile at the lower and upper boundaries of this zone, as illustrated in Fig.\ref{fig:eigen} for our model with Y$_c$ = 0.40. We recall that the internal structure and gradients of this model correspond to Fig.\ref{fig:grad} in the simpler case with XCO = 0\%. Another key observation comes from the associated spectrum in Fig.\ref{Fig:spectra-sem} in red. The local minima (red dots) correspond to the modes trapped in the semi-convective region, although most of these modes exhibit an exceptionally high energy density (Eq.\ref{eq:E}), as shown in Fig.\ref{Fig:inertia} with an order of magnitude $10^3$ times higher. 

{Their high energy density indicates that they are unlikely to be detected at the current level of data quality (see \cite{2014A&A...572A..11G} for a theoretical study of the detectability of the modes in red giants)}. Removing them from the spectrum and computing again the difference in consecutive modes period shows little to no effect on the period-spacing of the neighbouring modes. It suggests a strong decoupling between the cavities in the star, and thus making it quite difficult to detect the influence of these modes, even indirectly through their interactions with their neighbours. This point is consistent with previous studies \citep{Pincon2022}, suggesting that larger and sharper transitions at the top of the semi-convective region enhance the decoupling between oscillation cavities.

In the grid of models we studied, we can explore the visibility of modes trapped in the semi-convective region by examining less evolved models, particularly those near the emergence of this region around Y$_c$ = 0.56. In these cases, one or two trapped modes do not show such extreme jumps in their energies, possibly making them detectable in long-duration photometric observations. It is important to note that the visibility of these trapped modes could be influenced by other mixing prescriptions for the semi-convective zone or by varying mass regimes, presenting an interesting opportunity for further investigations. Indeed, a diversity of profiles are currently predicted by hydrodynamical simulations, while we have only tested here the classic induced semi-convection scheme used in other stellar evolution codes. Nevertheless, we show that from a theoretical point of view, mixed modes will be strongly influenced by the presence of a semi-convective zone, the main issue being the detectability of the modes trapped in these very deep layers.

\section{Conclusion} 

In this study, we examined the influence of internal mixing processes on seismic parameters of core helium-burning red giants throughout the sequence of evolution, focusing on how the \textit{helium-flash} discontinuity, or any physical processes responsible for this transition, as well as the overshooting parameter impact mixed-modes oscillation spectra. 
As a first step toward a further exploration of our models, we fixed the helium core mass $m_{He}$ to a fixed value of 0.50$\rm{M}_{\odot}$ to reduce the analysis complexity and isolate the effects of the other free parameters such as the effect of the \textit{helium-flash} remnants. We showed that:

\begin{itemize}[label=\textbullet]
    \item Sharp variations in chemical composition generate distinct mode-trapping effects, shaping the frequency structure and period-spacing patterns of oscillation modes.
    \item Variations in overshooting influence the extent of the semi-convective region, modulating the Brunt–Väisälä frequency and the observed period-spacing.
    \item Adjustments in the overshooting parameter are compensated by changes in the relative size of the semi-convective layer, leading to a relatively constant \textit{COS} zone size from the moment the semi-convective zone appears.
    \item While the model evolution remains mostly unaffected, the seismic signature is highly sensitive to these internal changes because of their strong impact on the Brunt–Väisälä frequency profile, which directly affects the period-spacing.
    \item Detailed asteroseismic studies are essential for characterising mixing processes in the mixed core region of core helium-burning stars.
\end{itemize}

The seismic signatures, notably seen in the Brunt–Väisälä frequency profile, show the potential of asteroseismic observations to probe the dynamics of these regions. Constraining the mixing properties at the boundary of the helium core will influence the amount of carbon and oxygen synthesised during core helium burning. Moreover, mixing has been shown to affect not only the properties of breathing pulses \citep{Castellani1985} but also the duration of the AGB phase and the properties of the AGB-bump \citep{Bossini2015, Dreau2022}. We refer the reader to \citet{Matteuzzi2025} for a detailed description of expected seismic signatures of different types of structural glitches in semi-analytical models of CHeB stars.

Recent 3D simulations \citep{Blouin2024, fuentes_3d_2025} have provided other perspectives on semi-convective mixing, with some studies emphasising the role of rotation in influencing semi-convective region. These results suggest that rotational effects could be an important factor to explore in future models. 

Future work will expand on this framework by exploring various mixing prescriptions, but also the treatment of mode coupling across multiple oscillation cavities \citep{Pincon2022,vrard_evidence_2022,Mosser2024Poster}. Starting from theoretical considerations and observations from a consistent model with various mixing prescriptions, we will compare the observed seismic signatures to our theoretical prediction and aim at building a consistent picture of the oscillation spectrum of low mass core He-burning stars, similarly to the work \cite{farnir_asteroseismology_2021} for subgiants. Developing detailed seismic analysis techniques to core helium-burning stars will yield valuable seismic constraints on mixing processes acting in stellar interiors and allow for a deeper characterisation of their chemical properties throughout this evolutionary phase. Ultimately, the derivation of seismic indicators, complemented by the potential integration in detailed modelling strategy, contributes to the broader objective of understanding chemical and angular momentum transport processes in stellar interiors through asteroseismology. Further investigation will also be carried out using evolutionary models to calibrate our mass of the helium core parameter $m_{He}$ more precisely, as well as the XCO parameter and the associated helium mass fraction profile between the convective core and the hydrogen shell.

{TESS and \textit{Kepler} have already observed more than 100.000 red giants across the sky, including numerous core-helium burning stars, providing a rich dataset for seismic analysis.} Additionally, the precise and accurate characterisation of core helium-burning stars is critical for the field of Galactic Archaeology, as it allows the study of more evolved stellar populations and provides a better understanding of the properties of convective cores during this unique evolutionary phase, which strongly influences chemical yields at later stages \citep{Herwig2005}. Applied in stellar clusters, a detailed characterisation of core He-burning stars would also be key to quantify the properties of mass loss on the RGB \citep{Howell2022,Tailo2022,Howell2024,Howell2025}, while detailed studies of the coupling properties of mixed modes in this evolutionary stage might help characterise non-standard products of evolution \citep{VanRossem2024}. In this context, the future PLATO mission \citep{2025ExA....59...26R} will play an important role for Galactic Archaeology \citep{Miglio2017}, while other missions such as HAYDN \citep{Miglio2021}, will provide the necessary data to test the limitations of our models in single stellar populations. {In this context, improving our asteroseismic analysis techniques is paramount to fully exploit the outputs of ongoing and future missions}.

\begin{acknowledgements}
We thank J. Montalbán for her contribution. L.P. is supported by the FNRS (''Fonds National de la Recherche Scientifique'') through a FRIA ''Fonds pour la Formation à la Recherche dans l’Industrie et l’Agriculture'') doctoral fellowship. G.B. acknowledges fundings from the Fonds National de la Recherche Scientifique (FNRS) as a postdoctoral researcher. M.M. and A.M. acknowledge support from the ERC Consolidator Grant funding scheme (project ASTEROCHRONOMETRY, \url{https://www.asterochronometry.eu}, G.A. n. 772293). 
\end{acknowledgements}

\bibliography{red}
\newpage
\begin{appendix}
\section{Effect of overshooting on oscillation spectrum}\label{A}
\begin{figure}[h]
\centering
    \includegraphics[width=9.5cm]{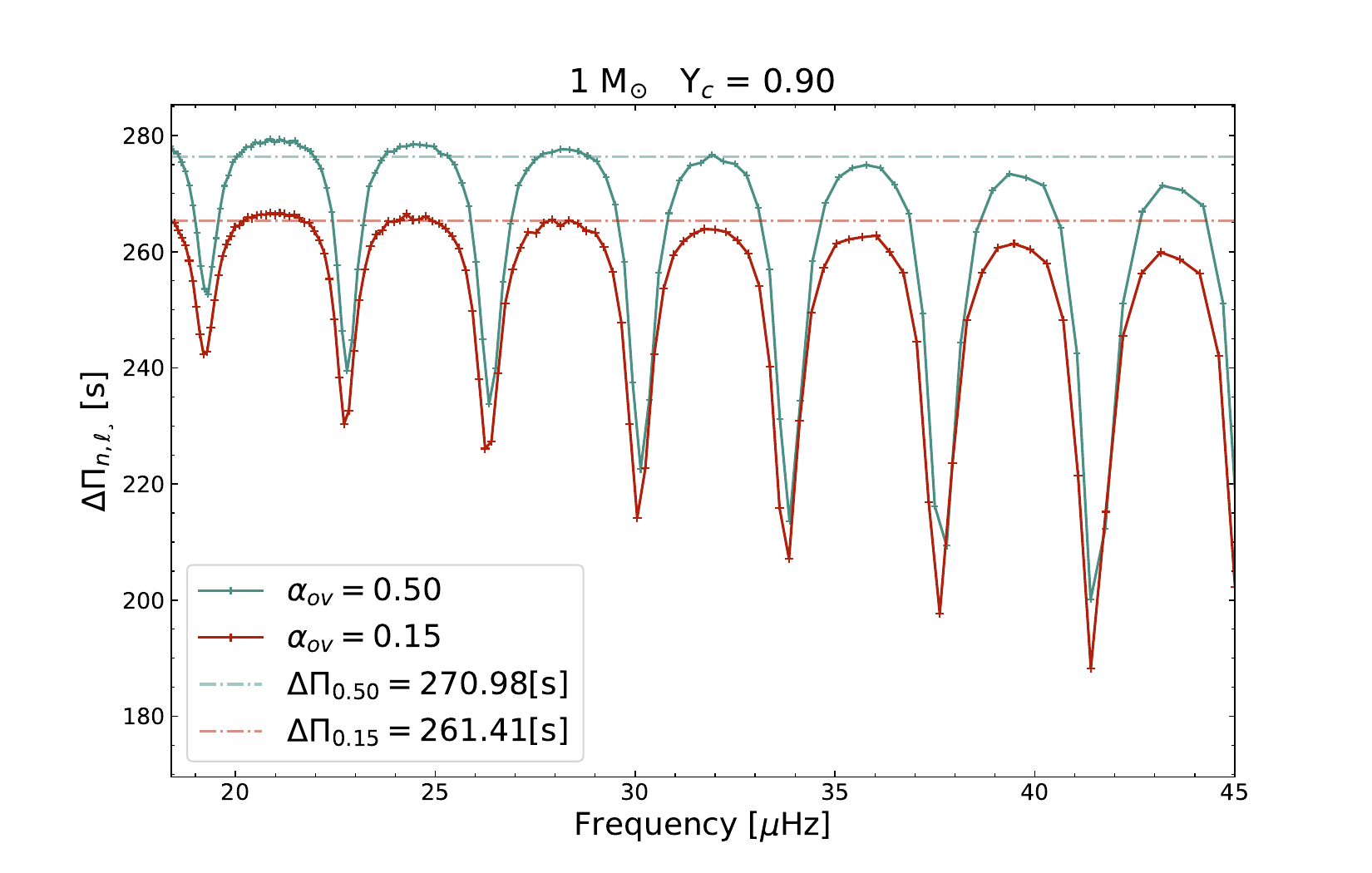}
\caption{Period-spacing of computed $\ell = 1$ mode frequencies, for models 1 M$_\odot$, Y$_c$ = 0.90 with two different value of overshooting parameter, $\alpha_{ov}$ = 0.50 (blue dots) and  $\alpha_{ov}$ = 0.15 (red dots), compared with their asymptotic value $\Delta \Pi$ (dashed lines).}
\label{Fig:spectra_over}
\end{figure}

As illustrated in Fig. \ref{Fig:spectra_over}, the individual period-spacing is influenced by the overshooting length put in the model. A model with a lower $\alpha_{ov}$ has a smaller fully mixed core and the integral of the Brunt-Väisälä frequency in the semi-convective layer is higher. This explains the lower asymptotic period-spacing in this case. 
\section{Kippenhahn diagrams}\label{B}
\begin{figure}[!h]
\begin{subfigure}{0.49\textwidth}
\includegraphics[width=9.5cm]{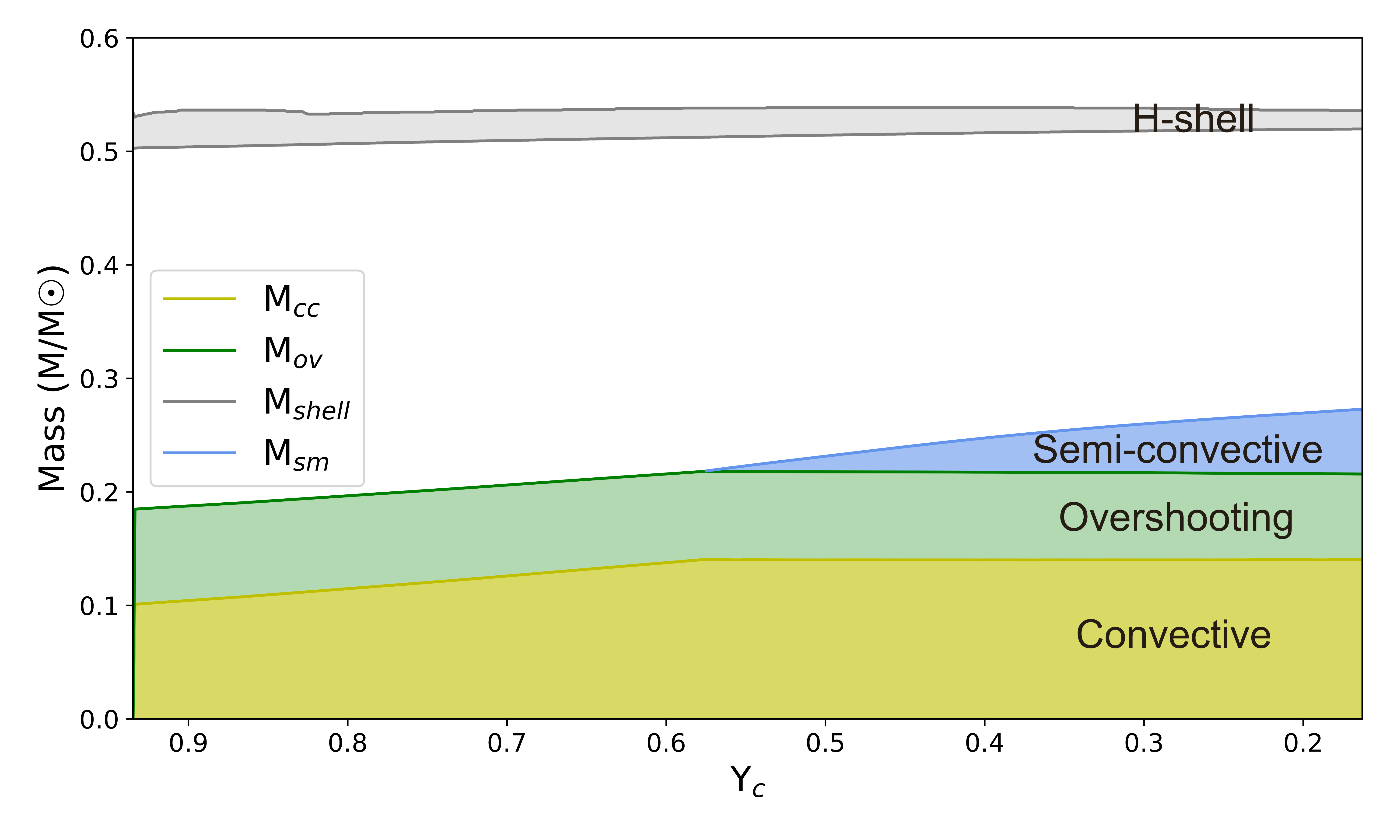}

\end{subfigure}
\begin{subfigure}{0.49\textwidth}
\centering
\includegraphics[width=9.5cm]{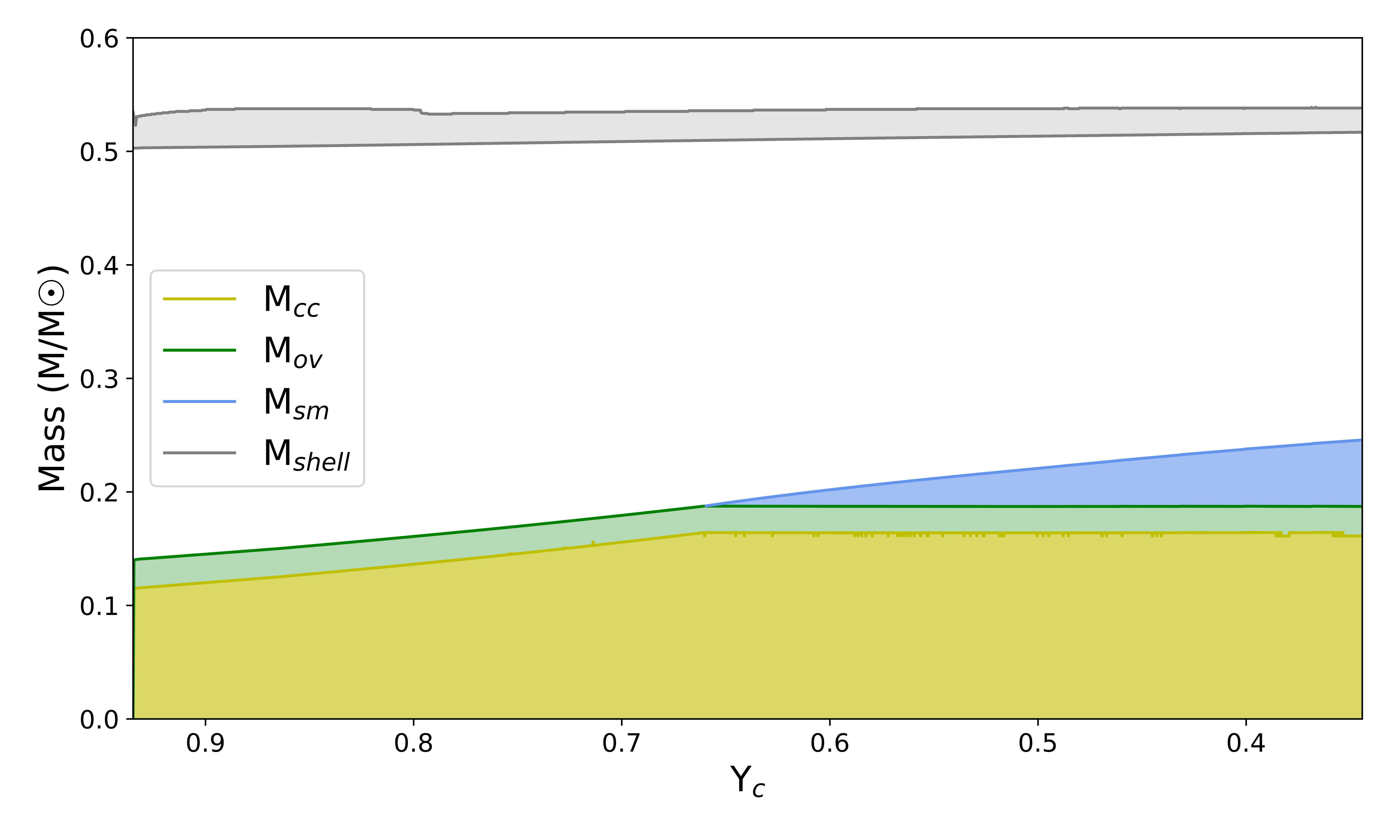}

\end{subfigure}
\caption{ Top panel: Kippenhahn diagram showing the evolution of the structure of the model 1M$_\odot$ with $\alpha_{ov}$=0.50 featuring the convective core, overshooting zone, semi-convective zone and hydrogen burning shell. Bottom panel: Kippenhahn diagram  showing the evolution of the structure of the model 1M$_\odot$ with $\alpha_{ov}$=0.15.}
\label{Fig:kippenhahn}
\end{figure}

{We show Kippenhahn diagrams (Fig. \ref{Fig:kippenhahn}) to illustrate the internal mixing processes during the core helium-burning phase of our models. Specifically, we present the models 1M$_\odot$ with $\alpha_{ov}$=0.50 and 0.15. These diagrams provide a representation of the evolution of the convective core, overshooting, and semi-convective regions within the star.}
\section{Energy density profile of an early model}\label{C}
\begin{figure}[h]
\centering
    \includegraphics[width=9.5cm]{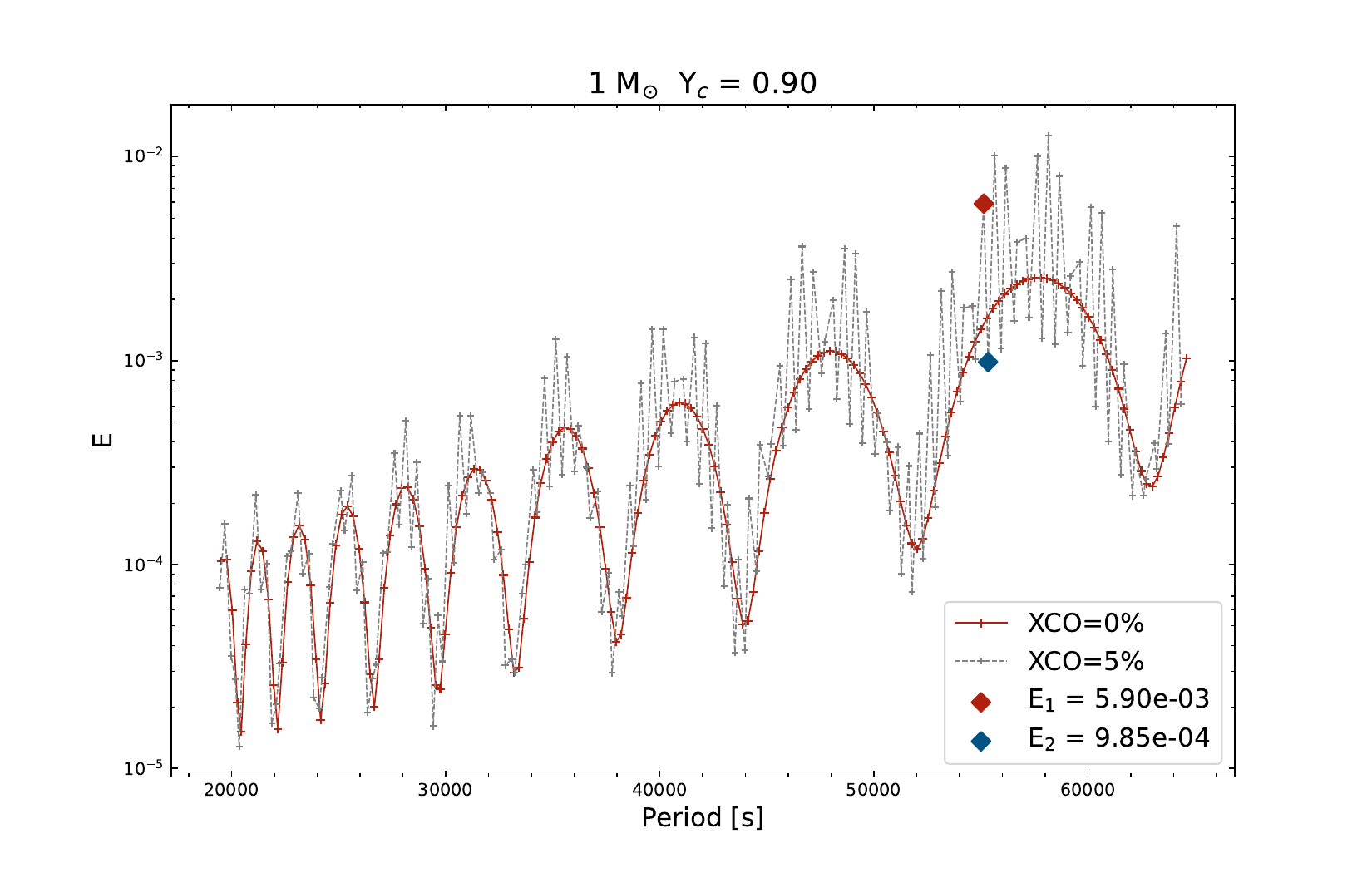}
\caption{Energy density of the modes of models 1 M$_\odot$, Y$_c$ = 0.90 for two sizes of the \textit{He-flash} discontinuity (implying changes in the density jump at this location) corresponding to fraction of carbon production of XCO = 5\% (grey dots) and XCO = 0\% (red dots). The red and blue markers indicate two selected modes.}
\label{Fig:inertia-90}
\end{figure}

The energy density profile for a $1\mathrm{M}\odot$ model with $Y_c = 0.90$ is shown for two cases: XCO 0\% and XCO = 5\%. The profiles display a sawtooth pattern corresponding to mode trapping caused by the discontinuity left by the helium flash, as illustrated in Fig. \ref{Fig:inertia-90}. The blue and red markers indicate two selected modes, one of which is trapped in the cavity formed by this discontinuity.
\end{appendix}
\end{document}